\documentclass[final,3p,times]{elsarticle}

\usepackage{lineno,hyperref,amsmath}

\usepackage{subfigure}
\usepackage{tabularx}

\journal{arXiv.org}

\begin{document}

\begin{frontmatter}

\title{A global nonhydrostatic dynamical core on cubed sphere using multi-moment finite volume method: formulation and preliminary test}

\author[XJTU]{Chungang Chen} 

\author[NWPC]{Xingliang Li}

\author[Titech]{Feng Xiao}

\author[NWPC]{Xueshun Shen\corref{correspondingauthor}}
\cortext[correspondingauthor]{Corresponding author}
\ead{shenxs@cma.gov.cn}

\address[XJTU]{State Key Laboratory for Strength and Vibration
of Mechanical Structures and School of Aerospace Engineering, Xi'an Jiaotong
University, Xi'an, China}
\address[NWPC]{Center of Numerical Weather Predication, China
Meteorological Administration, Beijing, China}
\address[Titech]{Department of Mechanical Engineering, Tokyo
Institute of Technology, Tokyo, Japan}

\begin{abstract}
A nonhydrostatic dynamical core has been developed by using the multi-moment finite volume method that ensures the rigorous numerical conservation. To represent the spherical geometry free of polar problems, the cubed-sphere grid is adopted. A fourth-order multi-moment discretization formulation is applied to the nonhydrostatic governing equations cast in local curvilinear coordinates on each patch of cubed sphere through a gnomonic projection. In vertical direction, the height-based terrain-following grid is used to represent the topography. To get around the CFL stability restriction imposed by relatively small grid spacing in the vertical direction, the dimensional-splitting time integration using the HEVI (Horizontal Explicit and Vertical Implicit) strategy is implemented by applying the IMEX Runge-Kutta scheme. The proposed dynamical core preserves the fourth-order accuracy in spherical geometry and has been verified by the widely-used benchmark tests. The results of our numerical experiments show that the present numerical core has superior solution quality and great practical potential as a platform for atmospheric models. A new unified model for numerical weather prediction and global atmospheric circulation simulation based on this dynamical core is under development.
\end{abstract}


\end{frontmatter}


\section{Introduction}

The multi-moment methods were proposed by introducing two or more kinds of quantities, which can be used to describe the spatial distribution of some physical field through point-wise values, volume (surface or line) integrated average, derivatives of different orders and so on, as model variables \cite{Xiao2004,Xiao2006} or constraints for spatial discretization \citep{ii2009}. With more local Degrees of Freedom (DOFs), the high-order multi-moment scheme has better flexibility in dealing with the different grid topologies and scalability on massive parallel clusters as other advanced schemes, like DG and SE schemes. Furthermore, the moments defined in our schemes have clear physical meanings and can result in the concise and flexible discretization procedures to be suited for the applications in different fields.

To develop the numerical models for atmospheric dynamics in spherical geometry, the computational meshes with quasi-uniform grid spacing, such as cubed-sphere grid, icosahedral geodesic grid and Yin-Yang grid, gain more and more attentions in the past decade with the applications of more powerful numerical schemes \cite{Staniforth2012}. The unified high-order formulations for global shallow water equations have been proposed on these three grids by using multi-moment finite volume method \citep{chen2014}. In this study, a fourth-order multi-moment finite volume formulations proposed in \citep{chen2008} are applied to constructed a nonhydrostatic dynamical core on cubed sphere. To extend the global shallow water model to 3D dynamical core, special attentions should be paid for developing the efficient time integration scheme, which can satisfy the operational requirement of numerical weather predication. In this study, we focus on how to get rid of the very restrictive CFL stability condition imposed by considerably small grid spacing in vertical direction for simulating the relative large-scale atmospheric dynamics. The horizontally-explicit and vertically-implicit (HEVI) strategy are adopted for time marching in this study with the application of implicit-explicit (IMEX) Runge-Kutta scheme.

The rest of this paper is organized as follows. In section 2, the numerical formulations of a multi-moment nonhydrostatic dynamical core are described in details. Some widely-used benchmark tests are checked to verify the proposed numerical model in section 3. And a short summary is finally given in section 4.

\section{Numerical formulations}

\subsection{Governing equations}

On each patch of cubed-sphere grid, the nonhydrostatic governing equations for atmospheric dynamics with shallow-atmosphere assumption are written in the flux-form as \cite{mcore2012,clark1977}
\begin{equation}
 \frac{\partial \boldsymbol{q}}                           {\partial t}
+\frac{\partial \boldsymbol{e}\left(\boldsymbol{q}\right)}{\partial \xi}
+\frac{\partial \boldsymbol{f}\left(\boldsymbol{q}\right)}{\partial \eta}
+\frac{\partial \boldsymbol{h}\left(\boldsymbol{q}\right)}{\partial \zeta}
=\boldsymbol{S}\left(\boldsymbol{q}\right),
\end{equation}
where $\left(\xi,\eta\right)$ are horizontal coordinates on each patch of the cubed sphere, $\zeta$ is a height-based terrain-following coordinate in vertical direction, $\boldsymbol{q}$ are dependent variables (predicted variables), $\boldsymbol{e}\left(\boldsymbol{q}\right)$, $\boldsymbol{f}\left(\boldsymbol{q}\right)$ and $\boldsymbol{h}\left(\boldsymbol{q}\right)$ are flux functions in $\xi$, $\eta$ and $\zeta$ directions, respectively  and $\boldsymbol{S}\left(\boldsymbol{q}\right)$ is source term.

The detailed expressions of governing equations used in this study is described with a brief introduction to the grid transformation as follows.

In the horizontal directions, the coordinates are $\xi=R\alpha$ and $\eta=R\beta$, where $R$ is radius of the Earth and $\alpha$, $\beta$ are central angles for a gnomonic projection varying within $\left[-\frac{\pi}{4},\frac{\pi}{4}\right]$ for each patch (details can be referred to \cite{chen2008}). In the vertical direction, $\zeta\in\left[0,r_t\right]$ is an uniform grid in the computational space, where $r_t$ is the model top. A non-uniform grid $\hat{\zeta}=\mathcal{T}\left(\zeta\right)$, which has smaller grid spacing near the surface, is adopted to better represent the surface topography and details are described in Appendix A.

The vertical mapping between the computational and the physical spaces is implemented through \cite{schar2002}
\begin{equation}
r=\hat{\zeta}+r_s\left(\xi,\eta\right)\frac{\sinh\left[\left(r_t-\hat{\zeta}\right)/S\right]}{\sinh\left(r_t/S\right)},\label{verticalmapping}
\end{equation}
where $r$ is the altitude, $r_s\left(\xi,\eta\right)$ is elevation of surface topography and the scale height $S=5000\mathrm{m}$ is adopted in this study.

The horizontal transformations between the longitude-latitude grid and the local curvilinear coordinates on each patch of cubed sphere are defined as follows.



The contravariant base vectors $\boldsymbol{a}^\xi$ and $\boldsymbol{a}^\eta$ are
\begin{equation}
\left\{
\begin{array}{c}
\boldsymbol{a}^\xi=\boldsymbol{i}\frac{1}{R\cos\theta}\frac{\partial \xi}{\partial \lambda}+\boldsymbol{j}\frac{1}{R}\frac{\partial \xi}{\partial \theta}\\
\boldsymbol{a}^\eta=\boldsymbol{i}\frac{1}{R\cos\theta}\frac{\partial \eta}{\partial \lambda}+\boldsymbol{j}\frac{1}{R}\frac{\partial \eta}{\partial \theta}\\
\end{array}
\right..
\end{equation}

The base vectors have the different expressions on different patches and can be derived from the projection relations.

%

The horizontal contravariant metric tensor is
\begin{equation}
\boldsymbol{G}^{ij}_H=\frac{\delta}{\left(1+X^2\right)\left(1+Y^2\right)}\left[\begin{array}{cc}1+Y^2&XY\\XY&1+X^2\end{array}\right],
\end{equation}
where $X=\tan{\alpha}$, $Y=\tan{\beta}$ and $\delta=\sqrt{1+X^2+Y^2}$.

The Jacobian of the horizontal transformation is
\begin{equation}
J_H=\left[\det\left({\boldsymbol{G}^{ij}_H}^{-1}\right)\right]^{\frac{1}{2}}.
\end{equation}

The contravariant velocity components are obtained by
\begin{equation}
\left\{
\begin{array}{c}
\tilde{u}=\boldsymbol{a}^\xi\cdot\boldsymbol{v}\\
\tilde{v}=\boldsymbol{a}^\eta\cdot\boldsymbol{v}\\
\end{array}
\right.,
\end{equation}
where $\boldsymbol{v}$ is the velocity vector in longitude-latitude coordinates.

The details of projection relations and transformation laws on cubed sphere can be referred to \cite{Nair2005a,Nair2005b,chen2008,mcore2012}.

In vertical direction, the governing equations in the height-based terrain-following coordinates can be derived through the chain rules \citep{clark1977}.

The Jacobian of vertical transform is $J_V=\frac{\partial r}{\partial \zeta}$, which can be directly obtained from Eq. \eqref{verticalmapping}. The components of contravariant metric tensor related with vertical transformation are $G_V^{13}=\frac{\partial \zeta}{\partial \xi}\left.\right|_{r=constant}$ and $G_V^{23}=\frac{\partial \zeta}{\partial \eta}\left.\right|_{r=constant}$. In the benchmark tests, these two components of contravariant metric tensor are analytically evaluated from the vertical mapping and the distribution of elevation of surface topography.

The overall Jacobian of transformation is written as $J=J_HJ_V$.

The dependent variables are
\begin{equation}
\boldsymbol{q}=\left[J\rho^\prime,J\rho \tilde{u}, J\rho \tilde{v}, J\rho w, J\left(\rho\theta\right)^\prime\right]^T,
\end{equation}
where $J$ is the Jacobian of the transformation, $\rho$ is density, $\tilde{u}$ and $\tilde{v}$ are contravariant velocity components in horizontal directions, $w$ is vertical velocity, $\theta$ is potential temperature and the superscript prime denotes the deviation with respect to the hydrostatic reference state.

In the atmospheric models, the thermodynamic variables are usually split into a reference state and the deviations to improve the accuracy of the simulation. We calculate the deviations in this study as

\begin{equation}
\left\{\begin{array}{l}
\rho^\prime\left(\xi,\eta,\zeta\right)=\rho\left(\xi,\eta,\zeta\right)-\overline{\rho}\left(\xi,\eta,\zeta\right)\\
\left(\rho\theta\right)^\prime\left(\xi,\eta,\zeta\right)=\rho\theta\left(\xi,\eta,\zeta\right)-\overline{\rho\theta}\left(\xi,\eta,\zeta\right)\\
\end{array}\right.,
\end{equation}
where the reference state satisfies the hydrostatic balance in vertical direction as
\begin{equation}
\frac{\partial \overline{p}\left(r\right)}{\partial r}=-g\overline{\rho}\left(r\right),
\end{equation}
and the deviation of pressure is
\begin{equation}
p^\prime\left(\xi,\eta,\zeta\right)=p\left(\xi,\eta,\zeta\right)-\overline{p}\left(\xi,\eta,\zeta\right).
\end{equation}



The flux functions are written in three directions as
\begin{equation}
\boldsymbol{e}=J\left[\rho\tilde{u},\rho\tilde{u}^2+G_H^{11}p^\prime,\rho\tilde{u}\tilde{v}+G_H^{11}p^\prime,\rho\tilde{u} w\right]^T,
\end{equation}
\begin{equation}
\boldsymbol{f}=J\left[\rho\tilde{v},\rho\tilde{u}\tilde{v}+G_H^{21}p^\prime,\rho\tilde{v}^2+G_H^{22}p^\prime,\rho\tilde{v} w\right]^T,
\end{equation}
and
\begin{equation}
\boldsymbol{h}=J\left[\rho\tilde{w},\rho\tilde{u}\tilde{w}+M^1p^\prime,\rho\tilde{v}\tilde{w}+M^2p^\prime,\rho\tilde{w}^2+ J_V^{-1}p^\prime\right]^T,
\end{equation}
where $\tilde{w}=\frac{1}{J_V}w+G^{13}_V\tilde{u}+G^{23}_V\tilde{v}$, $M^{s}=\left(G_V^{13}G_H^{s1}+G_V^{23}G_H^{s2}\right)$ ($s=1\ \mathrm{to}\ 2$).

The source term is written as
\begin{equation}
\boldsymbol{S}=\boldsymbol{S}_P+\boldsymbol{S}_C+\boldsymbol{S}_G+\boldsymbol{S}_R.
\end{equation}

$\boldsymbol{S}_P$ is the source term due to the grid transformation as
\begin{equation}
\boldsymbol{S}_P=\frac{2J}{R\delta^2}\left[0,AY\tilde{u},-BX\tilde{v},0,0\right]^T,
\end{equation}
and $\boldsymbol{S}_C$ is the source term representing the Coriolis force as
\begin{equation}
\boldsymbol{S}_C=\frac{2J\Omega}{\delta^2}\left[0,AY,BY,0,0\right]^T
\end{equation}
on patch one to four,
\begin{equation}
\boldsymbol{S}_C=\frac{2J\Omega}{\delta^2}\left[0,A,B,0,0\right]^T
\end{equation}
on patch five,
\begin{equation}
\boldsymbol{S}_C=-\frac{2J\Omega}{\delta^2}\left[0,A,B,0,0\right]^T,
\end{equation}
on patch six, where $\Omega$ is rotational speed of the Earth,
\begin{equation}
A=-XY\rho\tilde{u}+\left(1+Y^2\right)\rho\tilde{v}
\end{equation}
and
\begin{equation}
B=-\left(1+X^2\right)\rho\tilde{u}+XY\rho\tilde{v}.
\end{equation}

$\boldsymbol{S}_G$ is the source term for gravity force as
\begin{equation}
\boldsymbol{S}_G=\left[0,0,0,-Jg\rho^\prime,0\right]^T,
\end{equation}
where $g$ is gravitation constant.

$\boldsymbol{S}_R$ is the source term for Rayleigh friction
\begin{equation}
\boldsymbol{S}_R=\tau\left(\zeta\right)\rho\left[0,\tilde{u}-\tilde{u}_f,\tilde{v}-\tilde{v}_f,w-w_f,0\right]^T.
\end{equation}
where coefficient $tau\left(\zeta\right)$ determines the strength of Reyleigh friction, subscript $f$ indicates a reference velocity field.

\subsection{Definition of Degrees Of Freedom}

\begin{figure}[ht]
 \centering
 \includegraphics[width=0.45\textwidth]{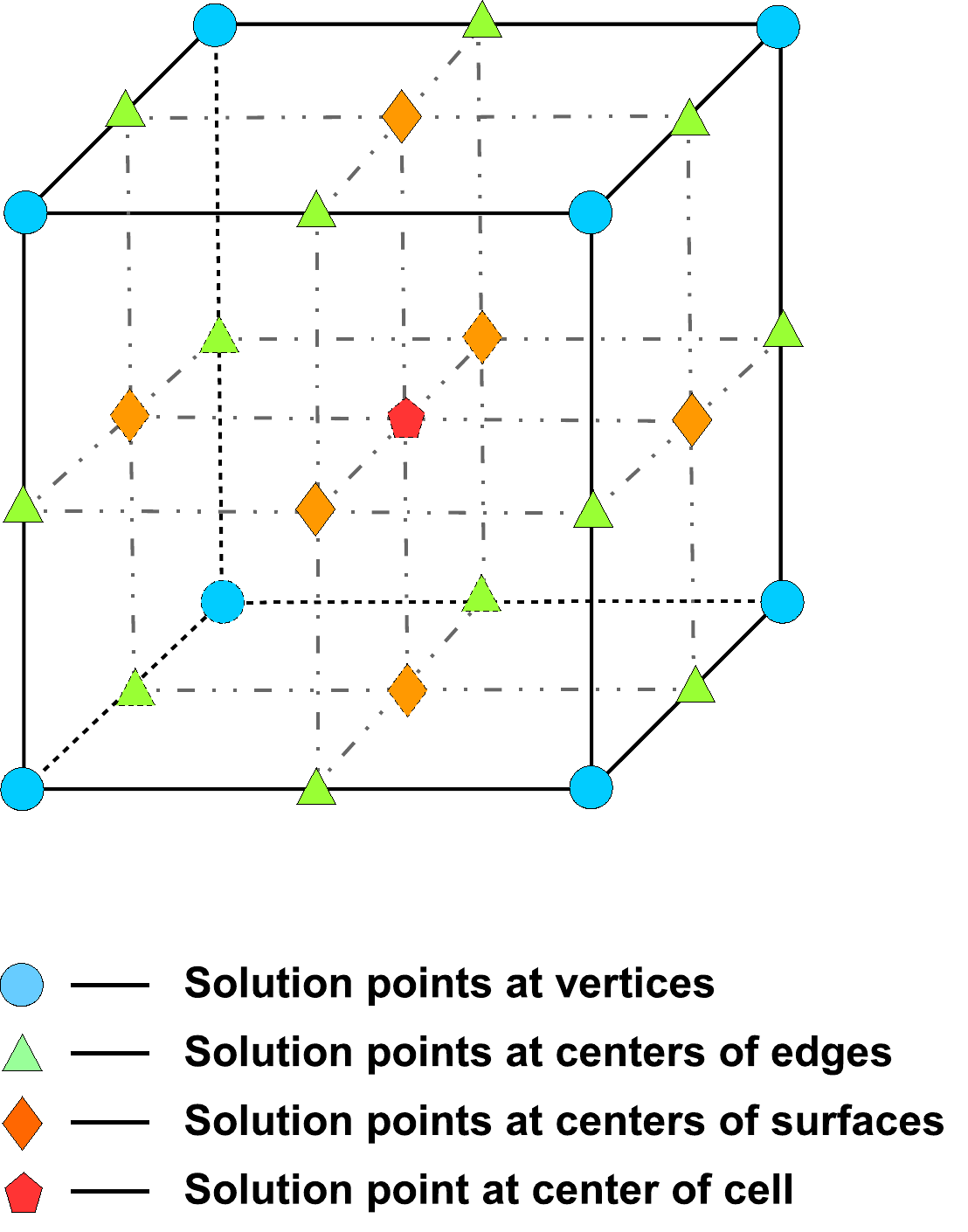}
  \caption{Definition of local DOFs within cell $\mathcal{C}^{ijkp}$.}\label{DOF}
\end{figure}

The multi-moment constrained finite volume (MCV) method \citep{ii2009} is adopted in this study. Twenty-seven point-wise values are defined as local DOFs (Degrees of Freedom) for each cell to construct the 3-point MCV scheme in three dimensions, as shown in Fig. \ref{DOF} for cell $\mathcal{C}^{ijkp}$, where superscripts $i,j,k$ denote the indices in $\xi$, $\eta$ ($i,j=1\ \mathrm{to}\ N_h$) and $\zeta$ ($k=1\ \mathrm{to}\ N_v$) directions and $p=1\ \mathrm{to}\ 6$ the number of the patch. The solution points are equidistantly distributed over the cell and the DOFs defined on the cell surfaces are shared by neighbouring cells.

The total number of computational cells adopted by the proposed model is $6{N_h}^2N_v$. The resolution in horizontal direction along the equator is $\frac{90^\circ}{N_h}$ in terms of computational cells and $\frac{45^\circ}{N_h}$ in terms of DOFs. In vertical directions, total number of layers is $2N_v+1$ including the surface and model top. Hereafter, we denote the computational mesh by its resolution $N_h\times N_v$.

\subsection{Spatial discretizations}

At solution points $P^{ijkp}_{mnl}$, where $m,n,l=1\ \mathrm{to}\ 3$ are local indices of DOFs within the computational cell  $\mathcal{C}^{ijkp}$, the local DOF are updated through a differential-form governing equations as
\begin{equation}
 \frac{\partial \boldsymbol{q}^{ijkp}_{mnl}}                           {\partial t}=
-\widehat{\boldsymbol{e}}_\xi\left(\xi_{im}\right)
-\widehat{\boldsymbol{f}}_\eta\left(\eta_{jn}\right)
-\widehat{\boldsymbol{h}}_\zeta\left(\zeta_{kl}\right)
+\boldsymbol{S}\left(\boldsymbol{q}^{ijkp}_{mnl}\right).\label{updateDOF}
\end{equation}

The MCV scheme in multi-dimensional case can be implemented by applying the one-dimensional formulations sweeping the different directions one-by-one \citep{ii2009}. Thus, we describe the numerical procedure of spatial discretization in $\xi$-direction as follows. Similar formulations can be derived for the spatial discretizations in $\eta$- and $\zeta$-directions. The details of multi-dimensional MCV discretization can be referred to \citep{ii2009}.

\begin{figure}[ht]
 \centering
 \begin{subfigure}[DOFs in 1D cell]
 {\includegraphics[scale=0.55]{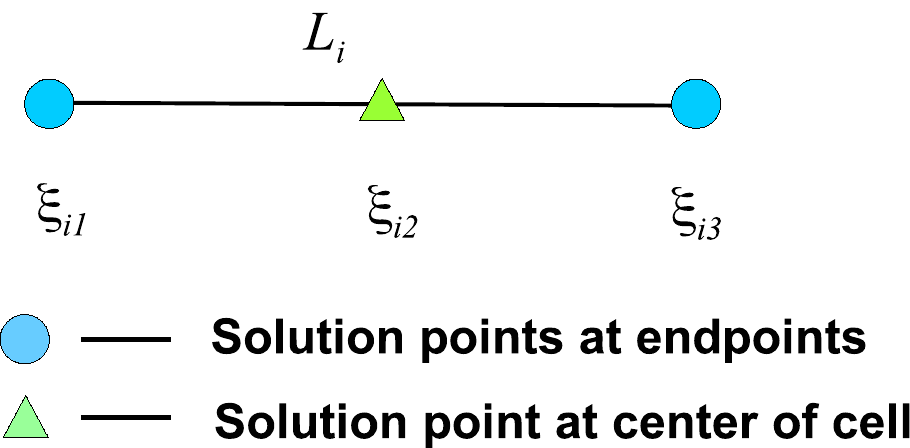}}
 \hspace{1cm}
 \end{subfigure}
 \begin{subfigure}[Updating DOF at cell center]
 {\includegraphics[scale=0.55]{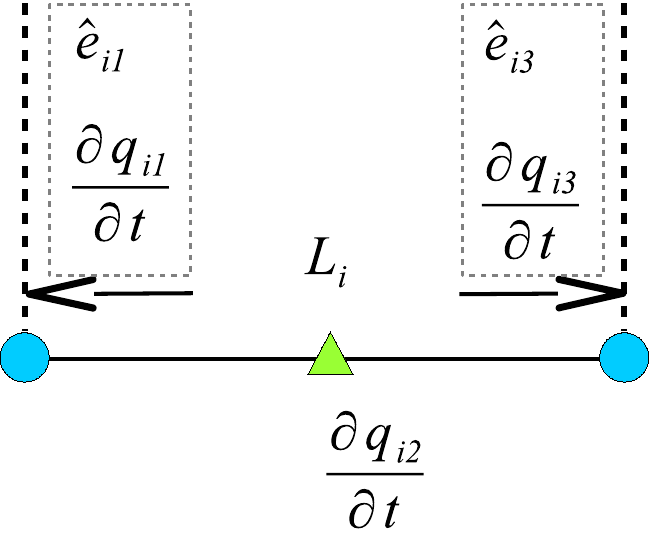}}
 \end{subfigure}

  \begin{subfigure}[Updating DOF at cell interface]
 {\includegraphics[scale=0.55]{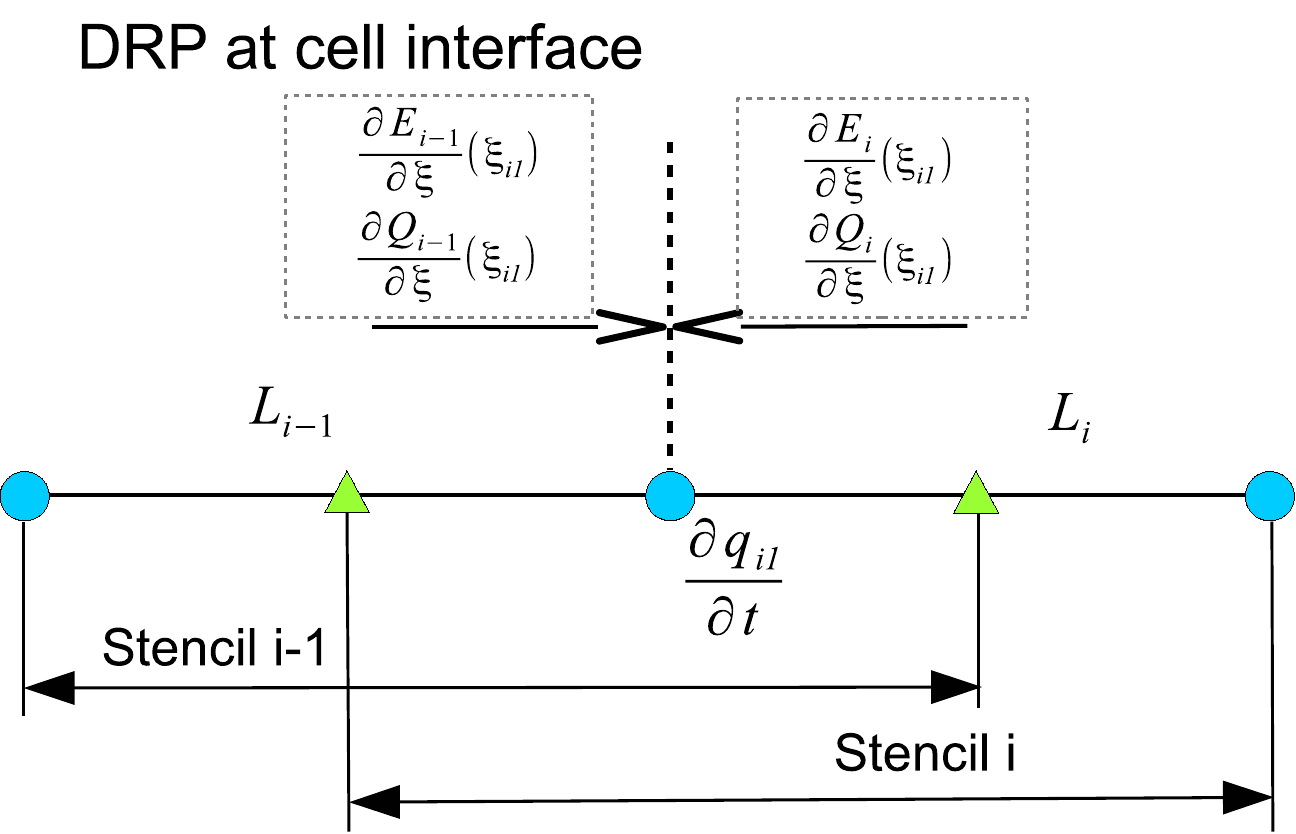}}
 \end{subfigure}
  \caption{Numerical scheme in one-dimensional case.}\label{1dFormulas}
\end{figure}

Considering the equations in one dimension as
\begin{equation}
\left(\frac{\partial \boldsymbol{q}}{\partial t}\right)^\xi+\frac{\partial \boldsymbol{e\left(q\right)}}{\partial \xi}=0,
\end{equation}
three local DOFs are defined within line segment $\mathcal{L}^{ijkp}_{nl}$ as shown in Fig. \ref{1dFormulas} (a) (one of 9 line segments along $\xi$-direction in Fig. \ref{DOF}), i.e., $\boldsymbol{q}^{ijkp}_{1nl}$, $\boldsymbol{q}^{ijkp}_{3nl}$ at cell interfaces (solid circles) and $\boldsymbol{q}^{ijkp}_{2nl}$ at cell center (solid triangle). Hereafter we use only the indices in $\xi$-direction for sake of brevity. As shown in Eq. \eqref{updateDOF}, the semi-discrete formulation for each DOF is written as
\begin{equation}
 \left(\frac{\partial \boldsymbol{q}_{im}}{\partial t}\right)^\xi =-\widehat{\boldsymbol{e}}_\xi\left(\xi_{im}\right).\label{1DupdateDOF}
\end{equation}

The different formulations are used to evaluated the derivatives of flux functions $\boldsymbol{e}$ at cell interfaces and center, as shown in Fig. \ref{1dFormulas} (b) and (c) respectively.

\begin{itemize}

\item Derivatives of flux functions $\boldsymbol{e}$ at cell interface ($\xi_{i1}$)

At interface, the derivatives of flux functions can be evaluated in two adjacent cells as shown in Fig. \ref{1dFormulas} (c). Generally, two different values are obtained. We solve a derivative Riemann problem here to derive an upwind scheme as
\begin{equation}
\widehat{\boldsymbol{e}}_\xi\left(\xi_{i1}\right)=\frac{1}{2}\left[\frac{\partial  \boldsymbol{E}_{i-1}}{\partial \xi}\left(\xi_{i1}\right)+\frac{\partial  \boldsymbol{E}_{i}}{\partial \xi}\left(\xi_{i1}\right)\right]+\frac{1}{2}\boldsymbol{a}_\xi\left[\frac{\partial  \boldsymbol{Q}_{i-1}}{\partial \xi}\left(\xi_{i1}\right)-\frac{\partial  \boldsymbol{Q}_{i}}{\partial \xi}\left(\xi_{i1}\right)\right],
\end{equation}
where $\boldsymbol{Q}$ and $\boldsymbol{E}$ are spatial reconstruction of predicted variables and flux functions, matrix $\boldsymbol{a}_\xi$ is determined by selected approximate Riemann solver in $\xi$-direction.

Using multi-moment concept, several interpolation profiles for spatial reconstruction has been developed \cite{chen2008,ii2009,chen2011,chen2015,BGS,WENO} for the schemes with different numerical properties. Considering the trade-off between the accuracy and the efficiency, the fourth-order profile developed in \citep{chen2008} is adopted in this study. The spatial reconstruction for line segment $L_{i-1}$ is a Lagrangian interpolation polynomial using four point-wise values of flux functions or predicted variables at $\xi_{i-1,1}$, $\xi_{i-1,2}$, $\xi_{i-1,3}$ and $\xi_{i2}$. Similarly, the point-wise values at $\xi_{i-1,2}$, $\xi_{i1}$, $\xi_{i2}$ and $\xi_{i3}$ are adopted for constraint conditions for spatial reconstruction within line segment $L_i$. The resulting multi-moment scheme is of fourth-order accuracy.

Three approximate Riemann solvers are investigated in \cite{Paul2010} in solving atmospheric dynamics. Considering the significance influence from the effects of the Coriolis force and the gravity force in atmospheric dynamics, specially for those large-scale atmospheric flows,  the waves propagate in a different way in comparison with the Euler equations for gas dynamics. The adopted Riemann solver should be carefully considered to accurately reproduce the wave propagation in atmosphere. A modified local Lax-Friedrichs (LLF) approximate Riemann solver is used in the proposed model for its simplicity. With the LLF solver, matrix $\boldsymbol{a}_\xi$ is simplified to be the maximal absolute value of eigenvalues of Jacobian matrix of flux functions $\frac{\partial \boldsymbol{e\left(\boldsymbol{q}\right)}}{\partial \boldsymbol{q}}$,which represents the maximal propagation speed related to the sound wave. In $\xi$-direction, it is written as
\begin{equation}
a_\xi=\left|\tilde{u}\right|+c_\xi,
\end{equation}
where the sound speed in the computational space is
\begin{equation}
c_\xi=\sqrt{G_H^{11}\gamma\frac{p}{\rho}}.
\end{equation}

In this study, the LLF solver is then modified by adopted a much smaller value of parameter $a$, which is specified as
\begin{equation}
a_\xi=\left|\tilde{u}\right|+K_u c_\xi,
\end{equation}
where $K_u$ is a parameter to adjust the effective of numerical viscosity, and $K_u=0.15$ is chosen in this study.

Since the physically-significant waves for large-scale atmospheric dynamics propagate much slower than the sound wave, this modification is expected to improve the accuracy of the proposed global model.

Analogously in $\eta$ and $\zeta$-directions, modified LLF solver is applied with $a_\eta=\left|\tilde{v}\right|+c_\eta$ and $a_\zeta=\left|\tilde{w}\right|+c_\zeta$, where the sound speeds in transformed coordinates
are
\begin{equation}
c_\eta=\sqrt{G_H^{22}\gamma\frac{p}{\rho}}
\end{equation}
and
\begin{equation}
c_\zeta=\sqrt{\left({J_V}^{-2}+M^1+M^2\right)\gamma\frac{p}{\rho}},
\end{equation}
respectively.
\item Derivatives of flux functions $\boldsymbol{e}$ at cell center ($\xi_{i2}$)

To guarantee the numerical conservation of the proposed model, the updating formulation of DOF at cell center is derived through the constraint condition on line-integrated average of the predicated variables
\begin{equation}
\overline{^{L_{\xi}}\boldsymbol{q}}_i=\frac{1}{\Delta \xi}\displaystyle\int_{\xi_{i1}}^{\xi_{i3}}\boldsymbol{q}\left(\xi\right)\mathrm{d}\xi,
\end{equation}

which can be approximated as
\begin{equation}
\overline{^{L_{\xi}}\boldsymbol{q}}_i=\frac{1}{6}\boldsymbol{q}_{i1}+\frac{2}{3}\boldsymbol{q}_{i2}+\frac{1}{6}\boldsymbol{q}_{i3}.\label{1DVIA}
\end{equation}
with above spatial reconstruction polynomial.

As a result, the updating formulation for DOF at cell center can be written as
\begin{equation}
\left(\frac{\partial \boldsymbol{q}_{i2}}{\partial t}\right)^\xi=\frac{3}{2}\left(\frac{\partial\overline{^{L_\xi}\boldsymbol{q}}_i}{\partial t}\right)^\xi-\frac{1}{4}\left[\left(\frac{\partial \boldsymbol{q}_{i1}}{\partial t}\right)^\xi+\left(\frac{\partial \boldsymbol{q}_{i3}}{\partial t}\right)^\xi\right],
\end{equation}
where the updating formulations of DOFs at cell interfaces have been obtained as above and the line-integrated average is updated with a flux-formulation as
\begin{equation}
\left(\frac{\partial\overline{^{L_\xi}\boldsymbol{q}}_i}{\partial t}\right)^\xi=-\frac{1}{\Delta \xi}\left(\hat{\boldsymbol{e}}_{i3}-\hat{\boldsymbol{e}}_{i1}\right)
\end{equation}
with the flux functions at cell interfaces estimated by DOFs defined at same locations directly.

The resulted scheme is conservative in terms of line-integrated average determined through Eq. \eqref{1DVIA}.

\end{itemize}

\subsection{Boundary condition}

In horizontal direction, one layer of ghost cells are supplemented for each patch. With enough ghost cells, the updating procedure is applied on each patch independently. The DOFs within ghost cells are evaluated by a single-cell based polynomial over the cell in adjacent patch. Furthermore, some DOFs, which are defined along the patch boundaries, can be updated in two or three patches and the different results are usually obtained during the simulation. A correction operation is applied by averaging the results from different patches. The construction of ghost cells in horizontal direction and the implementation of correction along the patch boundaries can be accomplished for a three-dimensional model by applying the numerical manipulation we have developed for the global shallow water model \cite{chen2008} at each model layer.

In vertical direction, the one-sided formulations are applied at surface and model top for spatial discretization in $\zeta$-direction. To evaluate the derivatives of flux functions $\boldsymbol{h}$, the formulations are implemented as (only the indices in $\zeta$-direction are showed here)
\begin{equation}
\widehat{\boldsymbol{h}}_\zeta\left(\zeta_{11}\right)=\frac{\partial  \boldsymbol{H}_{1}}{\partial \zeta}\left(\zeta_{11}\right)
\end{equation}
at surface and
\begin{equation}
\widehat{\boldsymbol{h}}_\zeta\left(\zeta_{N_v3}\right)=\frac{\partial  \boldsymbol{H}_{N_v}}{\partial \zeta}\left(\zeta_{N_v3}\right)
\end{equation}
at model top,
where the spatial reconstruction is accomplished through a quadratic Lagrangian interpolation based on three local DOFs within corresponding line segments to avoid introducing the ghost cells.

At bottom and top boundaries, the slip wall condition is applied by forcing $\tilde{w}=0$. Rayleigh friction is adopted in momentum equations near model top to assure the non-reflective boundary at model top in the tests with bottom topography and strength of Rayleigh friction is given as \cite{Durran1983}
\begin{equation}
\tau\left(\zeta\right)=
\left\{\begin{array}{ll}
0&\mathrm{if\ }\zeta<\zeta_D\\
\frac{\tau_0}{2}\left[1-\cos\left(\frac{\zeta-\zeta_D}{r_t-\zeta_D}\pi\right)\right] & \mathrm{if\ }0\leq\frac{\zeta-\zeta_D}{r_t-\zeta_D}\leq \frac{1}{2}\\
\frac{\tau_0}{2}\left[1+\sin\left(\frac{\zeta-\zeta_D}{r_t-\zeta_D}\pi-\frac{\pi}{2}\right)\right] & \mathrm{otherwise}
\end{array}\right.,
\end{equation}
where $\tau_0=-\frac{1}{6\times 3600}\ \mathrm{s}^{-1}$ and $\zeta_D=0.7r_t$.

In this study, the reference state of velocity field is chosen to be the initial condition.

\subsection{Time marching scheme}

Due to the very large ratio between the horizontal and the vertical grid spacings, the time step of the explicit time integration will be determined by the sound speed, the smallest grid spacing in vertical direction and the stability condition of the scheme. As a result, the available time step will has a magnitude less than one second in the practical applications with the veridical grid spacing of a few dozen meters near the surface. In this study, we use the implicit time integration to updating the terms related to the discretization in vertical direction and the stiff source terms including gravity force and Rayleigh friction. The implicit-explicit (IMEX) Runge-Kutta scheme is adopted to couple the explicit and implicit time marching. The time step is expected to be decided by the stability condition in horizontal direction, i.e. the horizonal velocity, sound speed and horizontal grid spacing. The time marching in the proposed model is accomplished from time step $n_t$ ($t=n_t\Delta t$) to $n_t+1$ as
\begin{equation}
\boldsymbol{q}^{n_t+1}=\boldsymbol{q}^{n_t}+\Delta t\sum_{r=0}^{R}\left[b_r\mathcal{H}\left(\boldsymbol{q}^{\left(r\right)}\right)+\tilde{b}_r\mathcal{V}\left(\boldsymbol{q}^{\left(r\right)}\right)\right],\label{imex}
\end{equation}
where
\begin{equation}
\boldsymbol{q}^{\left(r\right)}=\boldsymbol{q}^{n_t}+\Delta t\sum_{s=0}^{r-1}\left[a_{rs}\mathcal{H}\left(\boldsymbol{q}^{\left(s\right)}\right)\right]+\Delta t\sum_{s=0}^{r}\left[\tilde{a}_{rs}\mathcal{V}\left(\boldsymbol{q}^{\left(s\right)}\right)\right].
\end{equation}

At the $r^{th}$ substep, a nonlinear equation set, having the form of
\begin{equation}
\boldsymbol{y}\left(\boldsymbol{x}\right)=-\boldsymbol{x}+\boldsymbol{B}+\Delta t \tilde{a}_{rr}\mathcal{V}\left(\boldsymbol{x}\right)=0
\end{equation}
is solved to determine $\boldsymbol{q}^{\left(r\right)}$ by Newton's method, where
\begin{equation}
\boldsymbol{B}=\boldsymbol{q}^{n_t}++\Delta t\sum_{s=0}^{r-1}\left[a_{rs}\mathcal{H}\left(\boldsymbol{q}^{\left(s\right)}\right)+\tilde{a}_{rs}\mathcal{V}\left(\boldsymbol{q}^{\left(s\right)}\right)\right].
\end{equation}

The solution is approximately determined through iteration as
\begin{equation}
\left(\boldsymbol{I}-\Delta t \tilde{a}_{rr}\frac{\partial\mathcal{V}}{\partial \boldsymbol{x}}\left(\boldsymbol{x}_{z}\right)\right)\left(\boldsymbol{x}_{z+1}-\boldsymbol{x}_z\right)=\boldsymbol{y}\left(\boldsymbol{x}_z\right),
\end{equation}
which is solved using direct linear equation solver in this study with $\boldsymbol{x}_0=\boldsymbol{q}^{n_t }$.

The application of various IMEX Runge-Kutta scheme in the global atmospheric modelling were recently investigated in \cite{weller2013,Gardner2018}. In this study, a third-order, L-stable ARS343 ($R=3$ in Eq. \eqref{imex}) scheme proposed in \cite{ascher1997} is adopted. The coefficients adopted for Eq.\eqref{imex} are shown in Table \ref{IMEXcoef} for explicit part and Table \ref{IMEXcoef2} for implicit part.

\begin{table*}[ht]
{
\caption{Coefficients of explicit part of ARS343 scheme.}\label{IMEXcoef}
\begin{tabularx}{\textwidth}{l@{\extracolsep{\fill}}c|ccccc}

&     & 0            & 0            & 0            & 0            \\
&  a  & 0.4358665215 & 0            & 0            & 0            \\
&     & 0.3212788860 & 0.3966543747 & 0            & 0            \\
&     & -0.105858296 & 0.5529291479 & 0.5529291479 & 0            \\ \hline
& $b$ & 0            & 1.208496649  & -0.644363171 & 0.4358665215  \\
\end{tabularx}}
\end{table*}

\begin{table*}[ht]
{
\caption{Coefficients of implicit part of ARS343 scheme.}\label{IMEXcoef2}
\begin{tabularx}{\textwidth}{l@{\extracolsep{\fill}}c|ccccc}
&             & 0 & 0            & 0            & 0            \\
& $\tilde{a}$ & 0 & 0.4358665215 & 0            & 0            \\
&             & 0 & 0.2820667392 & 0.4358665215 & 0            \\
&             & 0 & 1.208496649  & -0.644363171 & 0.4358665215 \\ \hline
& $\tilde{b}$ & 0 & 1.208496649  & -0.644363171 & 0.4358665215 \\
\end{tabularx}}
\end{table*}

As $\tilde{b}=\tilde{a}_{Rs}$ ($s=0\ \mathrm{to}\ 3$) in ARS343 scheme, with $\boldsymbol{q}^{\left(R\right)}$ evaluated at $R^{th}$ substep the numerical results at next time step is obtained by
\begin{equation}
\boldsymbol{q}^{n_t+1}=\boldsymbol{q}^{n_t}+\Delta t\sum_{r=0}^{R}\left[\left(b_r-a_{Rr}\right)\mathcal{H}\left(\boldsymbol{q}^{\left(r\right)}\right)\right].
\end{equation}

\section{Numerical results}

\subsection{Convergence test}

\begin{figure}[h]
	\centering
	\includegraphics[width=0.5\textwidth]{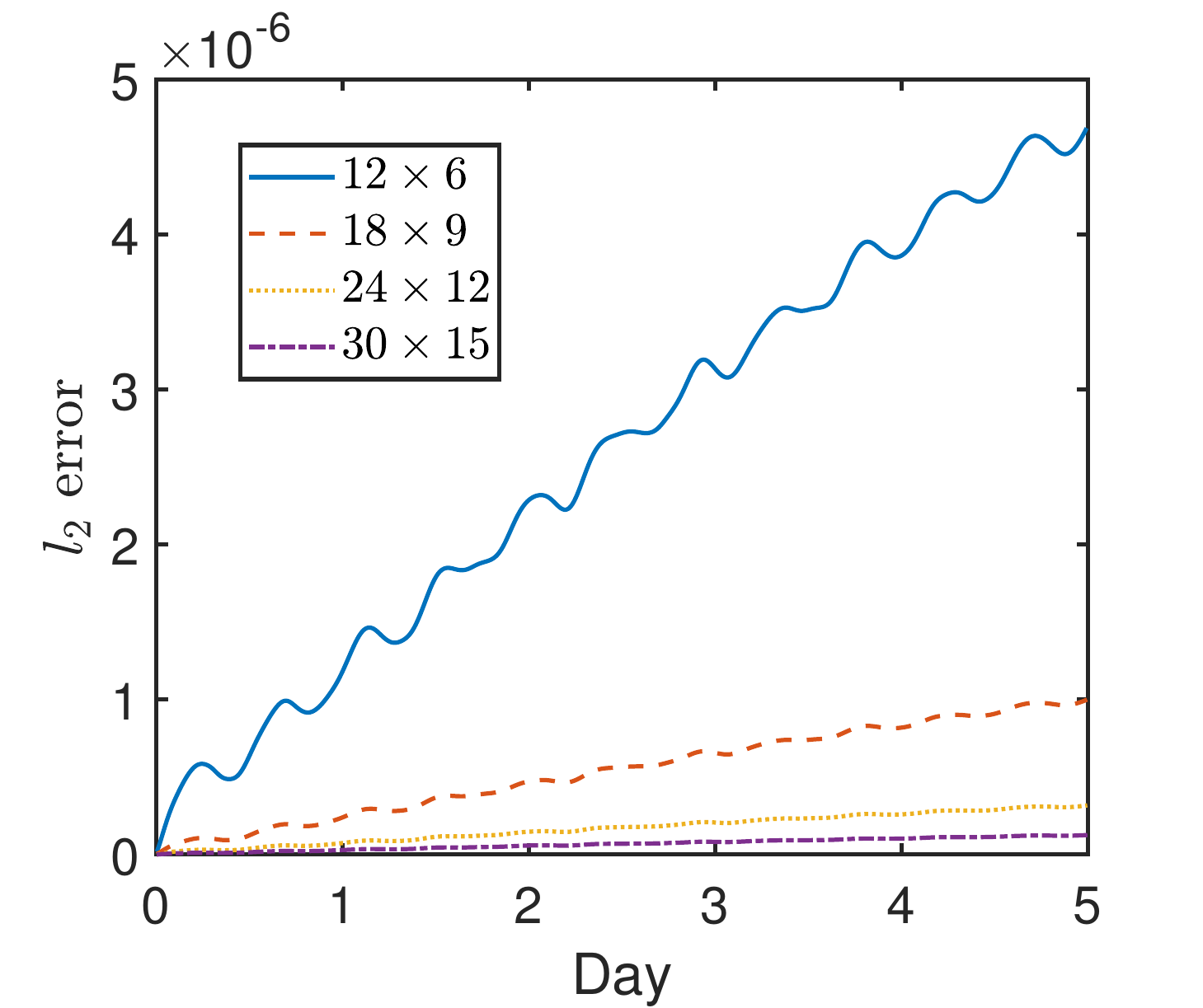}
	\caption{Time history of normalized $l_2$ errors of air density on a series of refining grids.}\label{errorhistory}
\end{figure}

The convergence rate of the proposed model is first checked. The initial condition is specified same as mountain-induced Rossby wave case \citep{jablonowski2008}. Excluding the bottom mountain here, this balanced condition will be preserved during the simulation. As a result, the normalized errors can be calculated based on the difference between the numerical solution and initial condition. The time history of normalized $l_2$ errors of air density on a series of refining grids are shown in Fig. \ref{errorhistory}. The normalized errors at day 5 and corresponding convergence rate are shown in Table \ref{convergencerate}. The fourth-order convergence rate is well preserved for 3D global model.

\begin{table*}[ht]
{
\caption{Normalized $l_2$ errors and convergence rates on a series of refining grids at day 5.}\label{convergencerate}
\begin{tabularx}{\textwidth}{l@{\extracolsep{\fill}}ccccc}
\hline\hline
&  Resolution        & time steps &  $l_2$ error &  Convergence rate    \\ \hline
& $12 \times 6$    &   576   &    $4.6851\times10^{-6}$  &   -            \\
& $18 \times 9$    &   864   &    $9.9927\times10^{-7}$  &  3.8107   \\
& $24 \times 12$  & 1152   &    $3.1751\times10^{-7}$  &  3.9853   \\
& $30 \times 15$  & 1440   &    $1.2612\times10^{-7}$  &  4.1375    \\ \hline
\end{tabularx}}
\end{table*}
\subsection{Results of DCMIP 2008 cases}
Some benchmark tests proposed in \citep{jablonowski2008} are then checked.  The horizontal grid resolutions are $1^\circ$ and $1.5^\circ$ in different tests and the corresponding time steps are specified as 200s and 300s respectively to satisfy the CFL stability condition in horizontal directions. Uniform vertical grid is utilized for two cases without bottom mountain. The model top is 30km for 3D Rossby-Haurwitz wave test and 10km for gravity wave test. Non-uniform vertical grid is adopted to better represent the effect of topography in other two tests, which is described in details in section \ref{verticalgrid}.

\begin{figure}[h]
 \centering
 \begin{subfigure}[850hPa zonal velocity]
  { \includegraphics[width=0.48\textwidth]{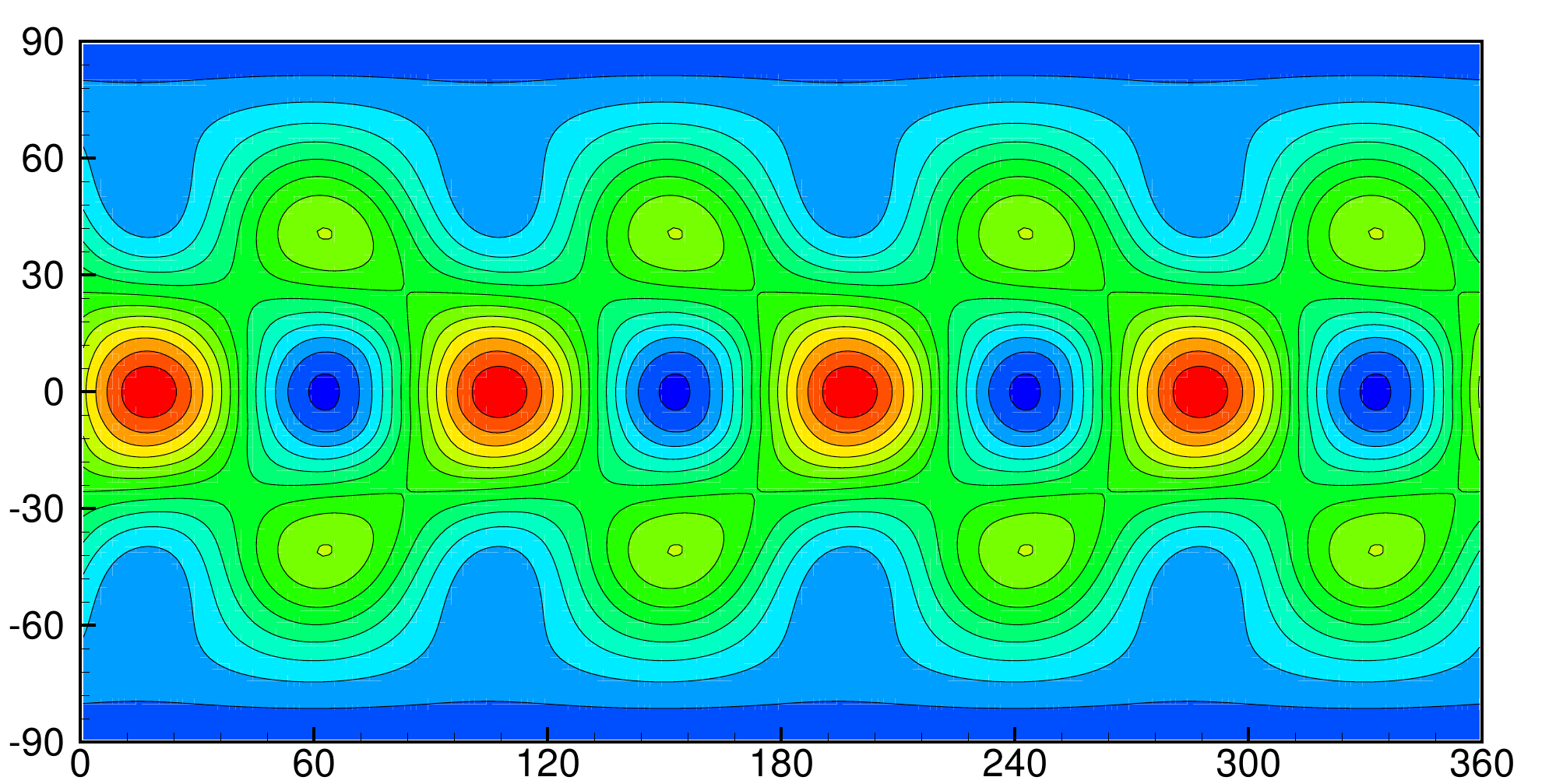}}
  \end{subfigure}
   \begin{subfigure}[850hPa meridional velocity]
   { \includegraphics[width=0.48\textwidth]{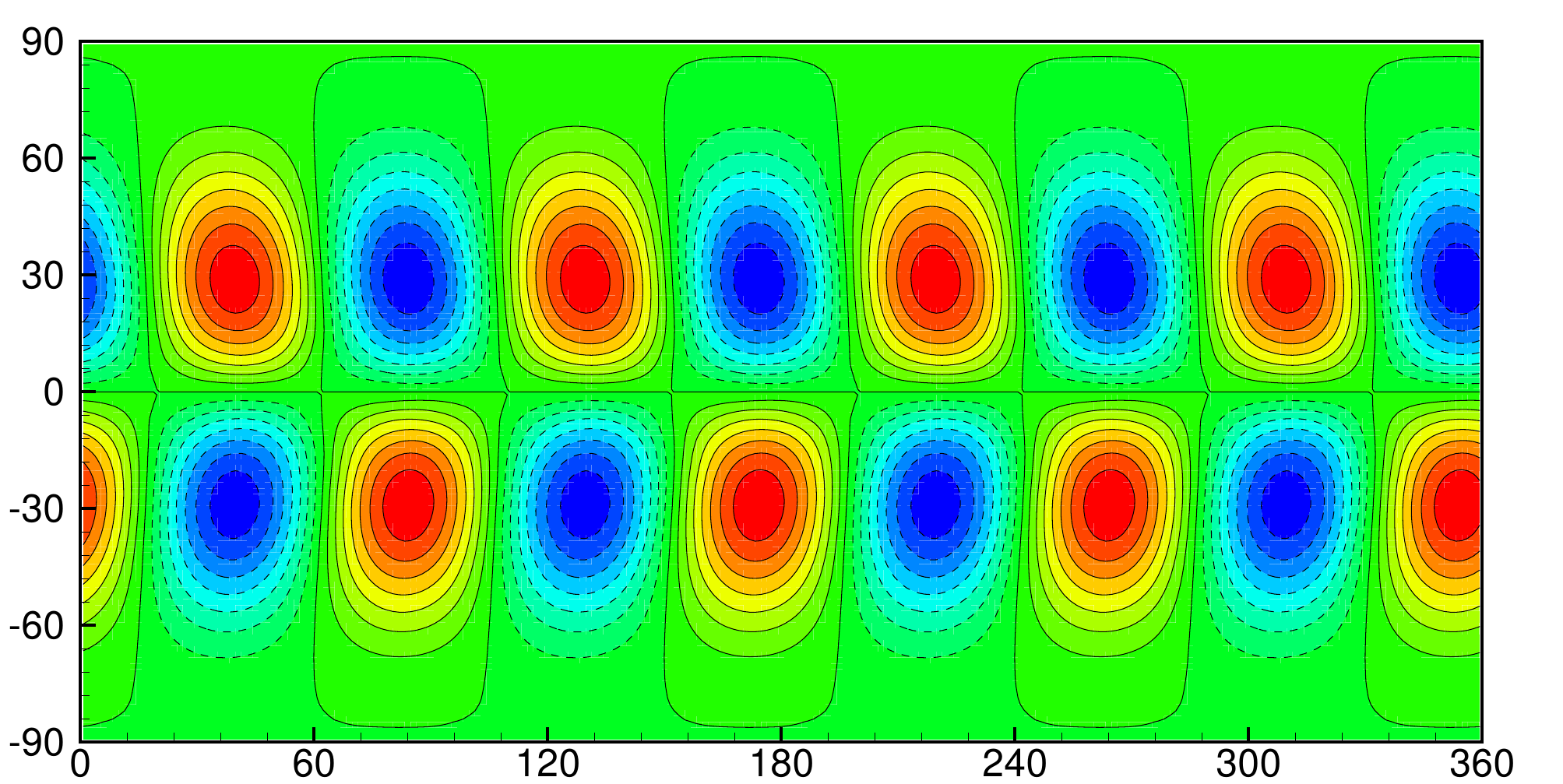}}
   \end{subfigure}
\begin{subfigure}[Surface pressure]
  { \includegraphics[width=0.48\textwidth]{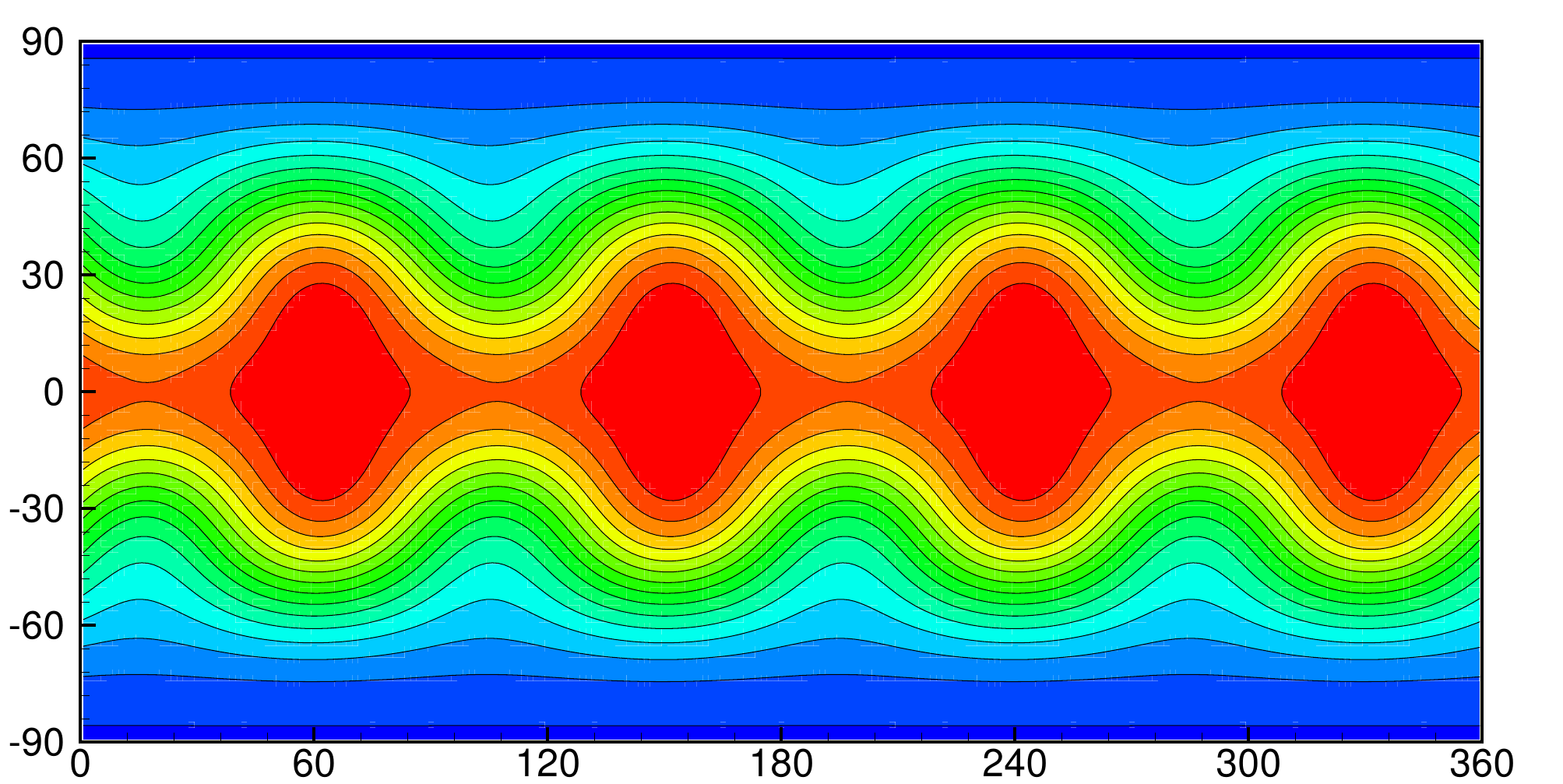}}
  \end{subfigure}
   \begin{subfigure}[850hPa temperature]
   { \includegraphics[width=0.48\textwidth]{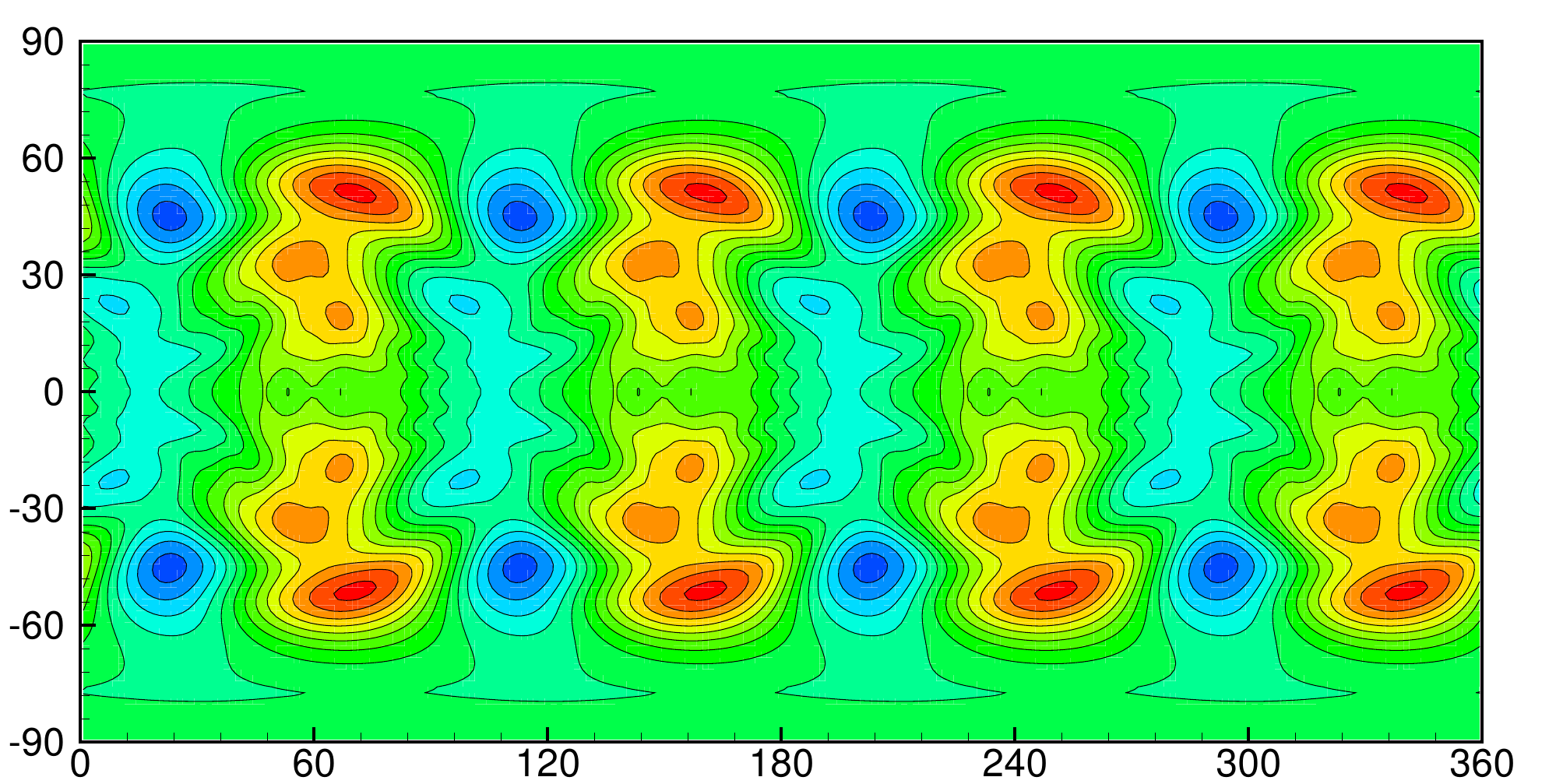}}
   \end{subfigure}
   \begin{subfigure}[500hPa height]
  { \includegraphics[width=0.48\textwidth]{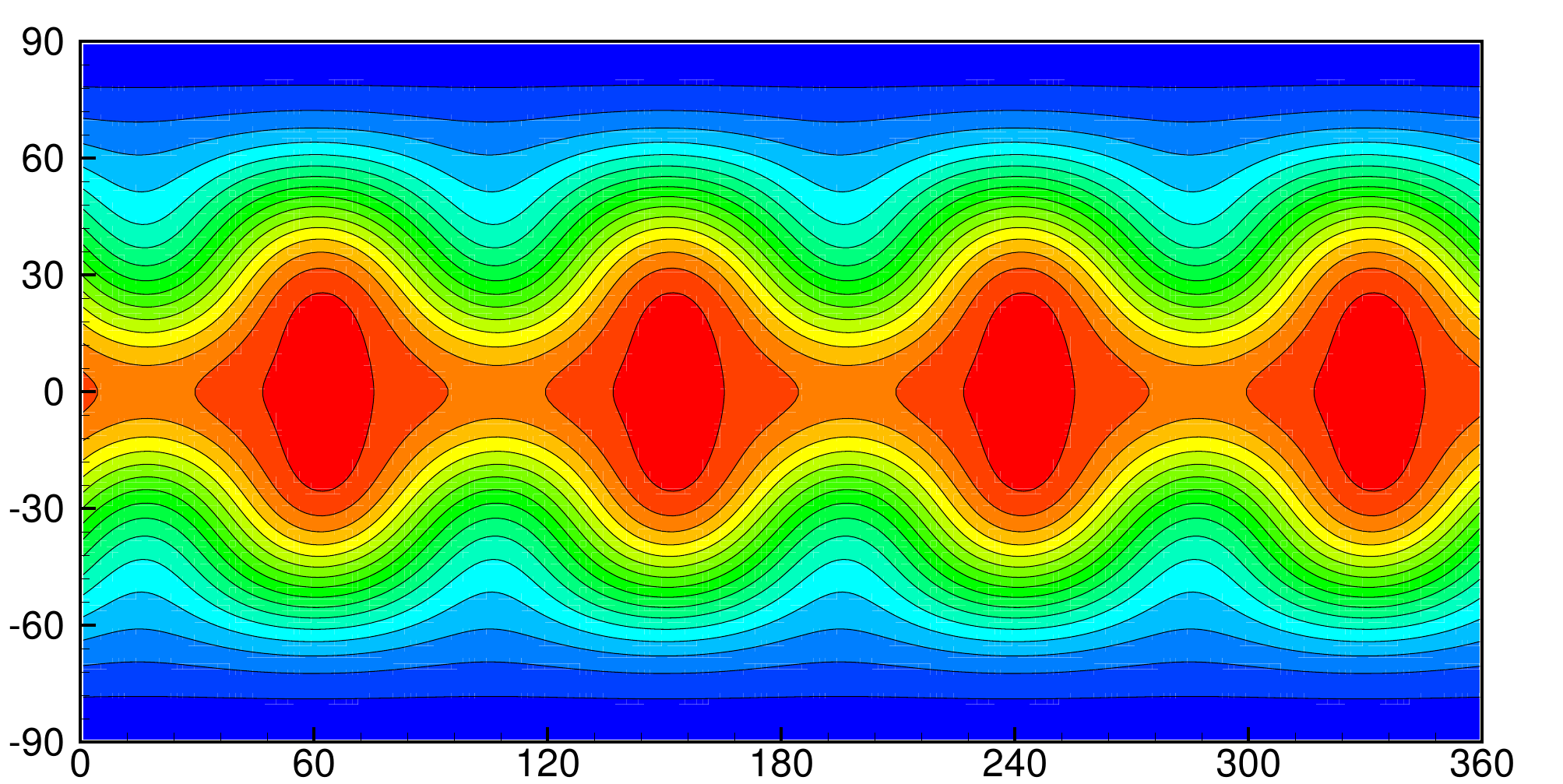}}
  \end{subfigure}
   \begin{subfigure}[850hPa vertical velocity]
   { \includegraphics[width=0.48\textwidth]{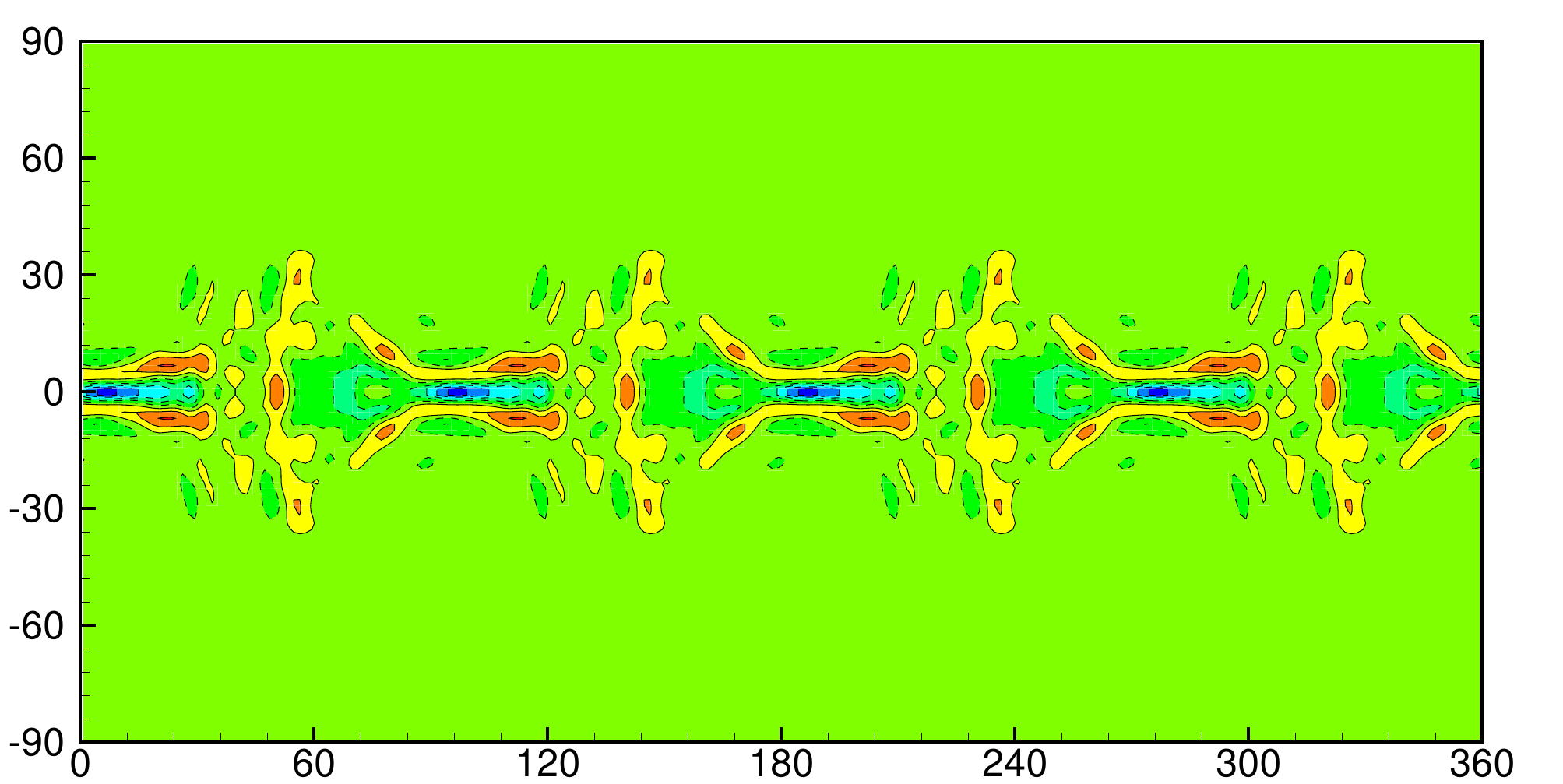}}
   \end{subfigure}
  \caption{Contour plots of numerical results of 3D Rossby-Haurwitz wave on grid $30 \times 15$ at day 15. Displayed contour lines vary within $\left[0\mathrm{m/s},24\mathrm{m/s}\right]$ with an interval of 2m/s for 850hPa zonal velocity, within $\left[-14\mathrm{m/s},14\mathrm{m/s}\right]$ with an interval of 2m/s for 850hPa meridional velocity, within $\left[5160\mathrm{m},5760\mathrm{m}\right]$ with an interval of 40m for 500hPa height, within $\left[955\mathrm{hPa},1025\mathrm{hPa}\right]$ with an interval of 5hPa for surface pressure and within $\left[281.48\mathrm{K},282\mathrm{k}\right]$ with an interval of 0.04k for 850hPa temperature and within $\left[-0.0018\mathrm{m/s},0.001\mathrm{m/s}\right]$ with an interval of 0.0004m/s for 850hPa vertical velocity. The dashed lines are used for negative values.}\label{RossbyWave}
\end{figure}

\begin{figure}[h]
 \centering
 \begin{subfigure}[Hour 6]
  { \includegraphics[width=0.48\textwidth]{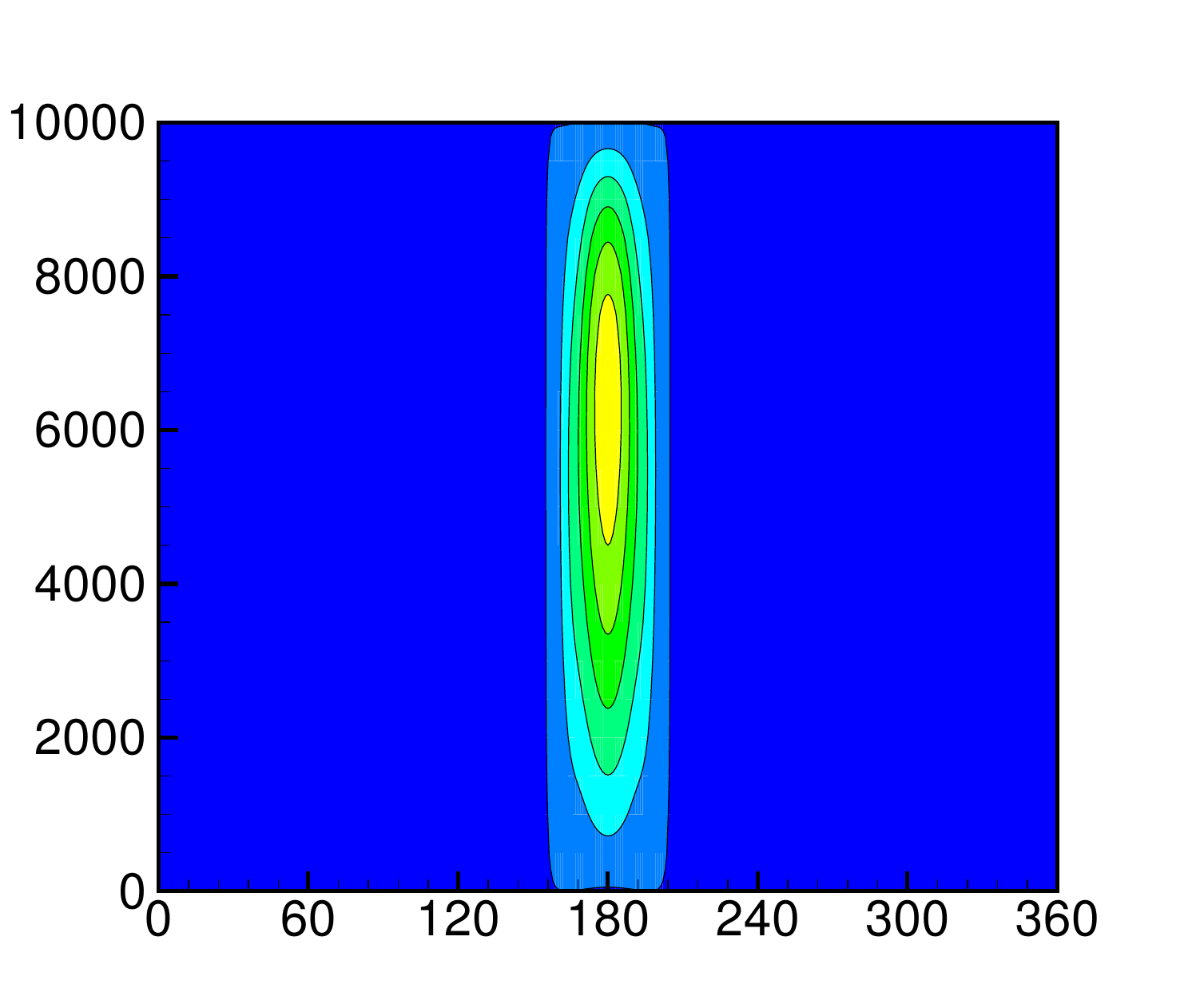}}
  \end{subfigure}
   \begin{subfigure}[Hour 12]
   { \includegraphics[width=0.48\textwidth]{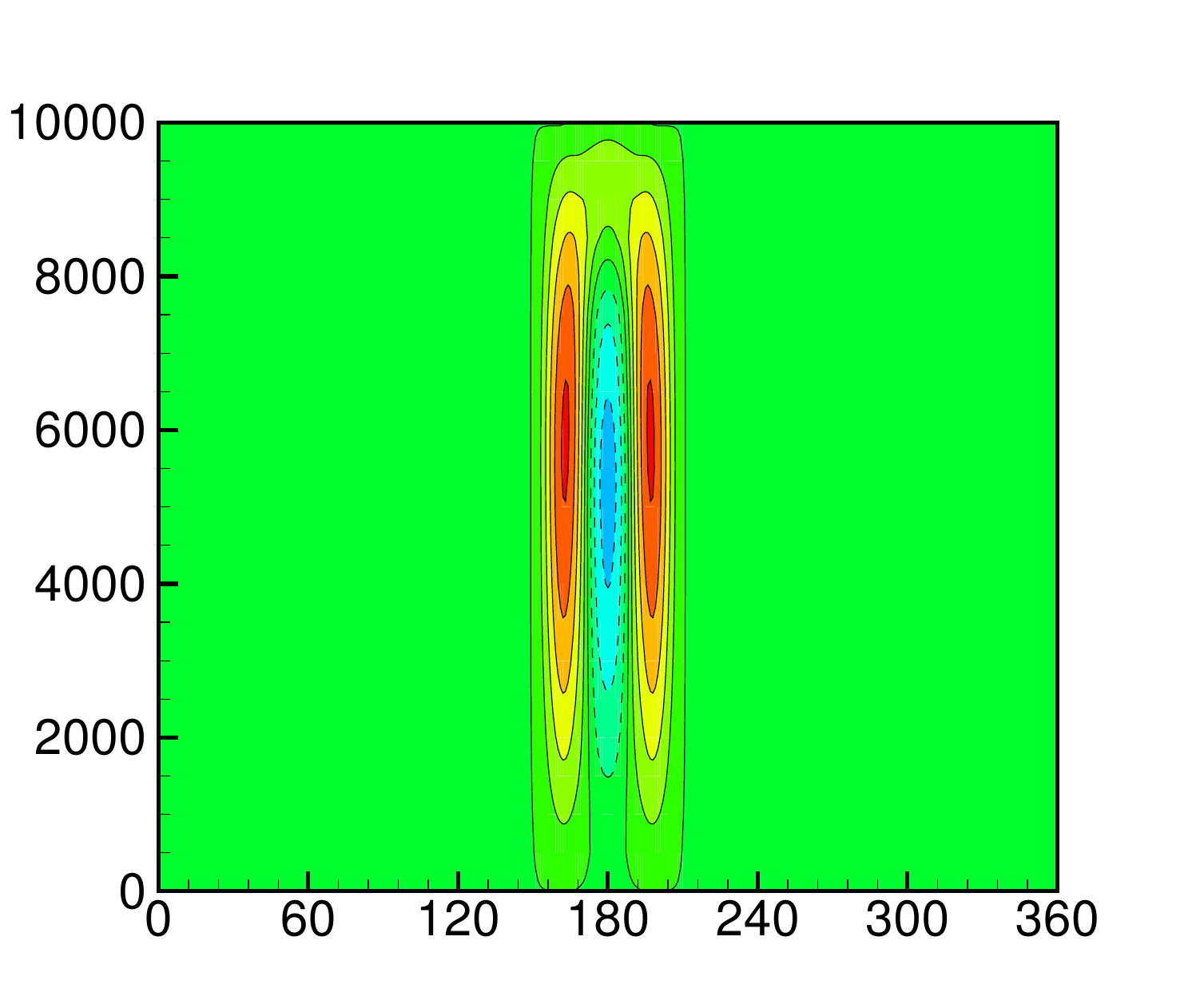}}
   \end{subfigure}
\begin{subfigure}[Hour 24]
  { \includegraphics[width=0.48\textwidth]{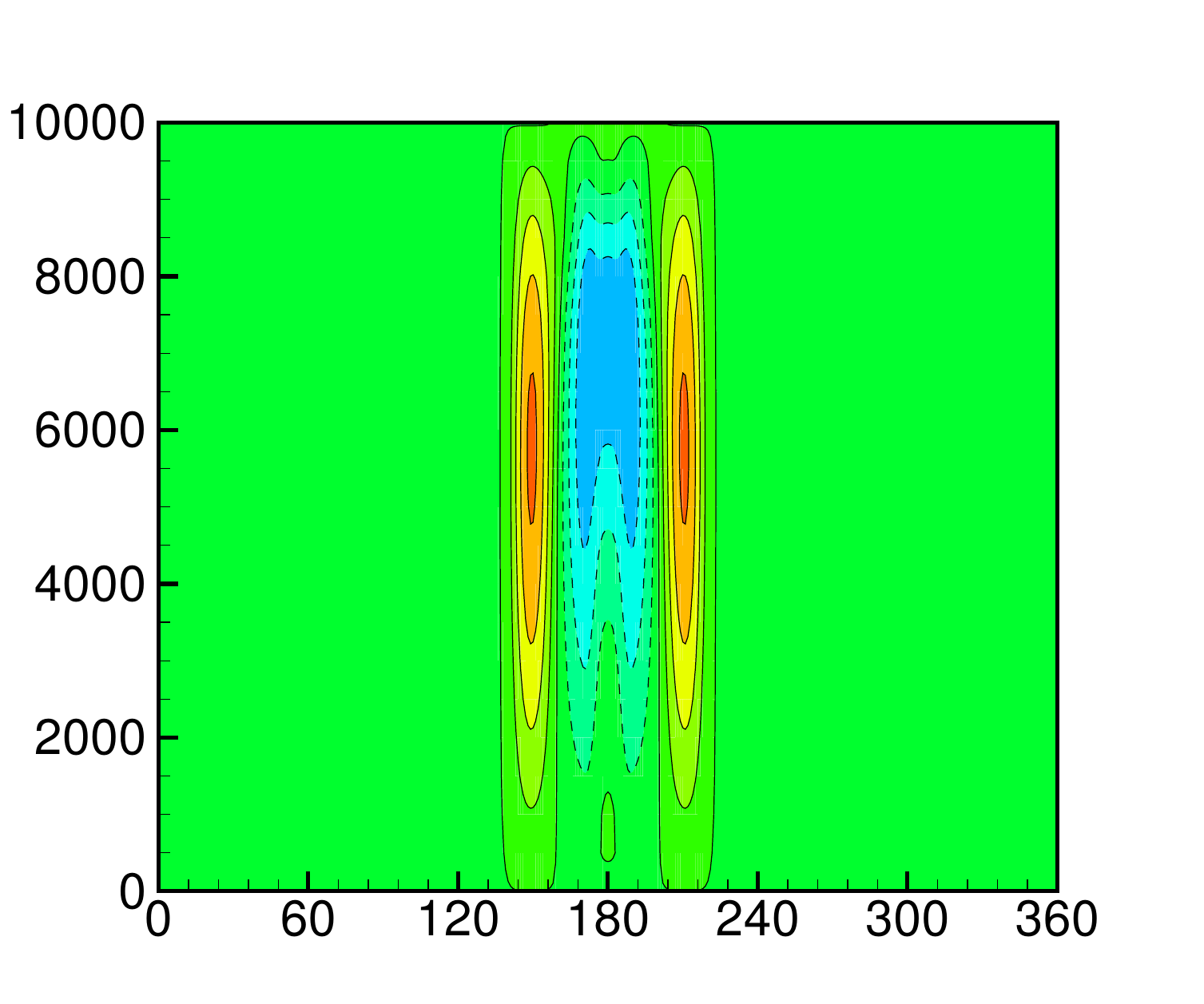}}
  \end{subfigure}
   \begin{subfigure}[Hour 48]
   { \includegraphics[width=0.48\textwidth]{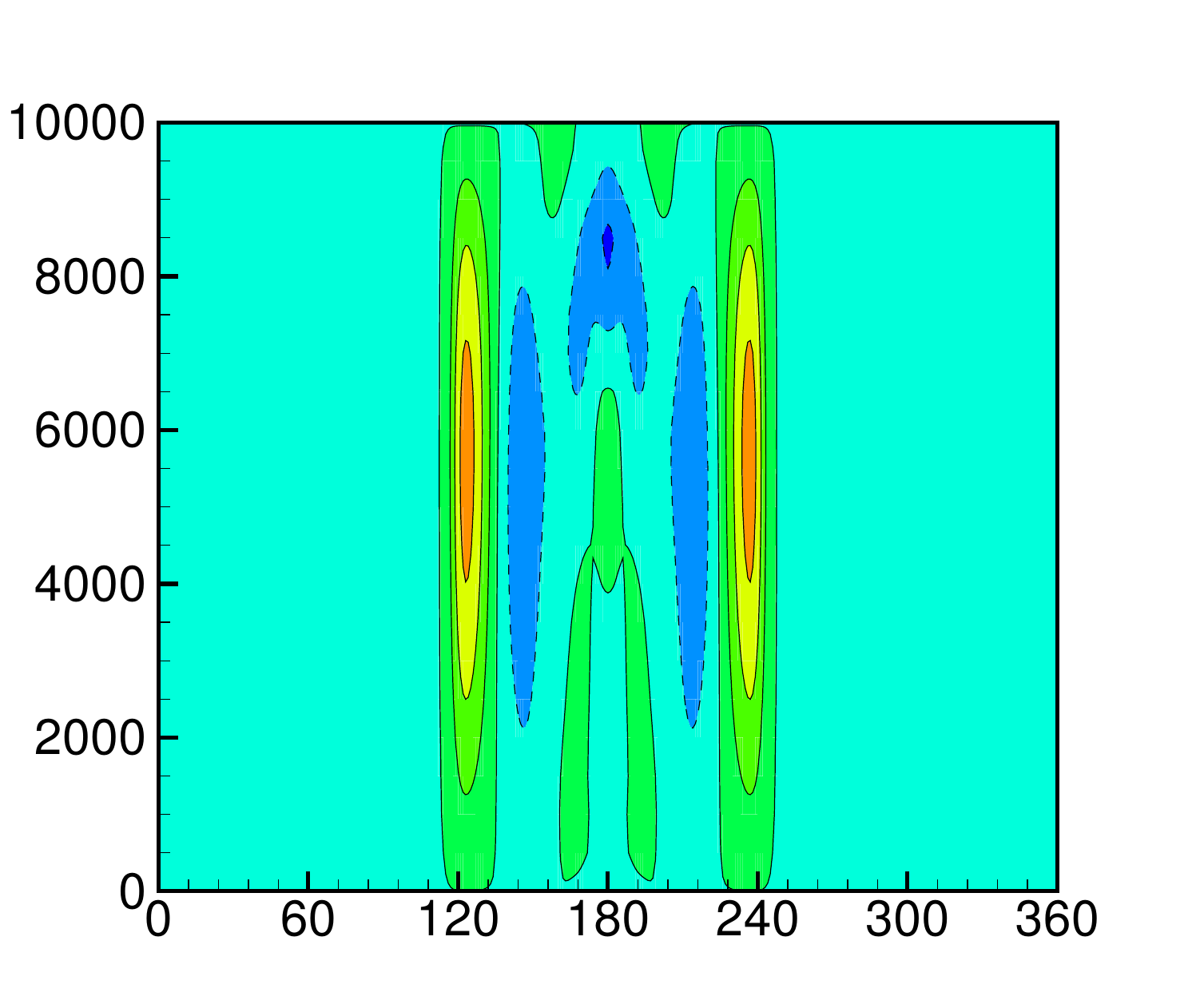}}
   \end{subfigure}
   \begin{subfigure}[Hour 72]
  { \includegraphics[width=0.48\textwidth]{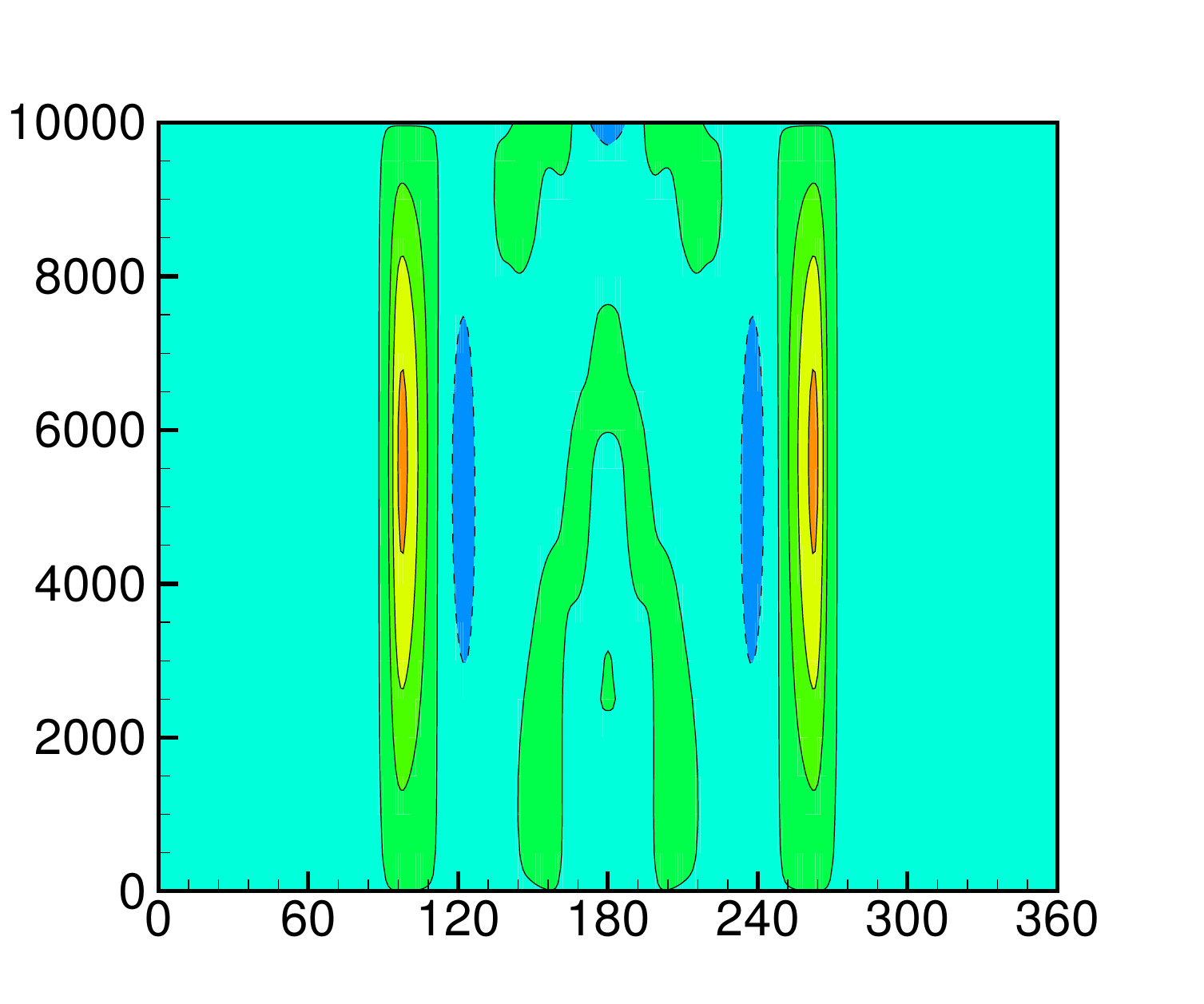}}
  \end{subfigure}
   \begin{subfigure}[Hour 96]
   { \includegraphics[width=0.48\textwidth]{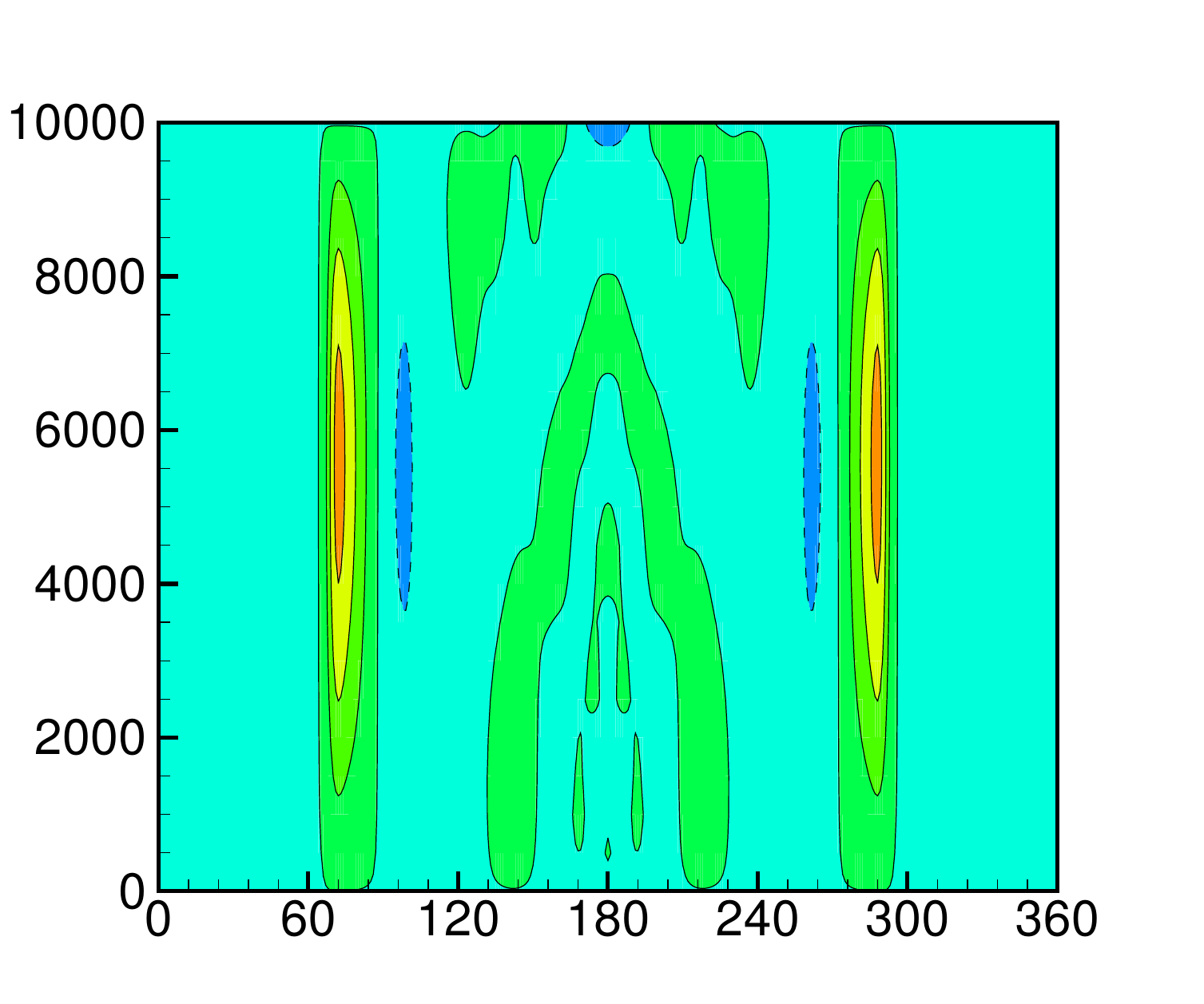}}
   \end{subfigure}
  \caption{Contour plots of numerical results of gravity wave tests on grid $45 \times 10$, shown are potential temperature perturbation. Displayed contour lines vary within $\left[0\mathrm{K},7\mathrm{K}\right]$ with an interval 1K at hour 6, within $\left[-2.5\mathrm{K},2.5\mathrm{K}\right]$ with an interval 0.5K at hour 12 and 24 and within $\left[-1\mathrm{K},2\mathrm{K}\right]$ with an interval 0.5K at hour 48, 72 and 96. The contour line of 0.01K is displayed here instead of 0K in all plots. The dashed lines are used for negative values.}\label{GravityWave}
\end{figure}

\begin{figure}[h]
 \centering
 \begin{subfigure}[700hPa geopotential height at day 5]
  {\includegraphics[width=0.48\textwidth]{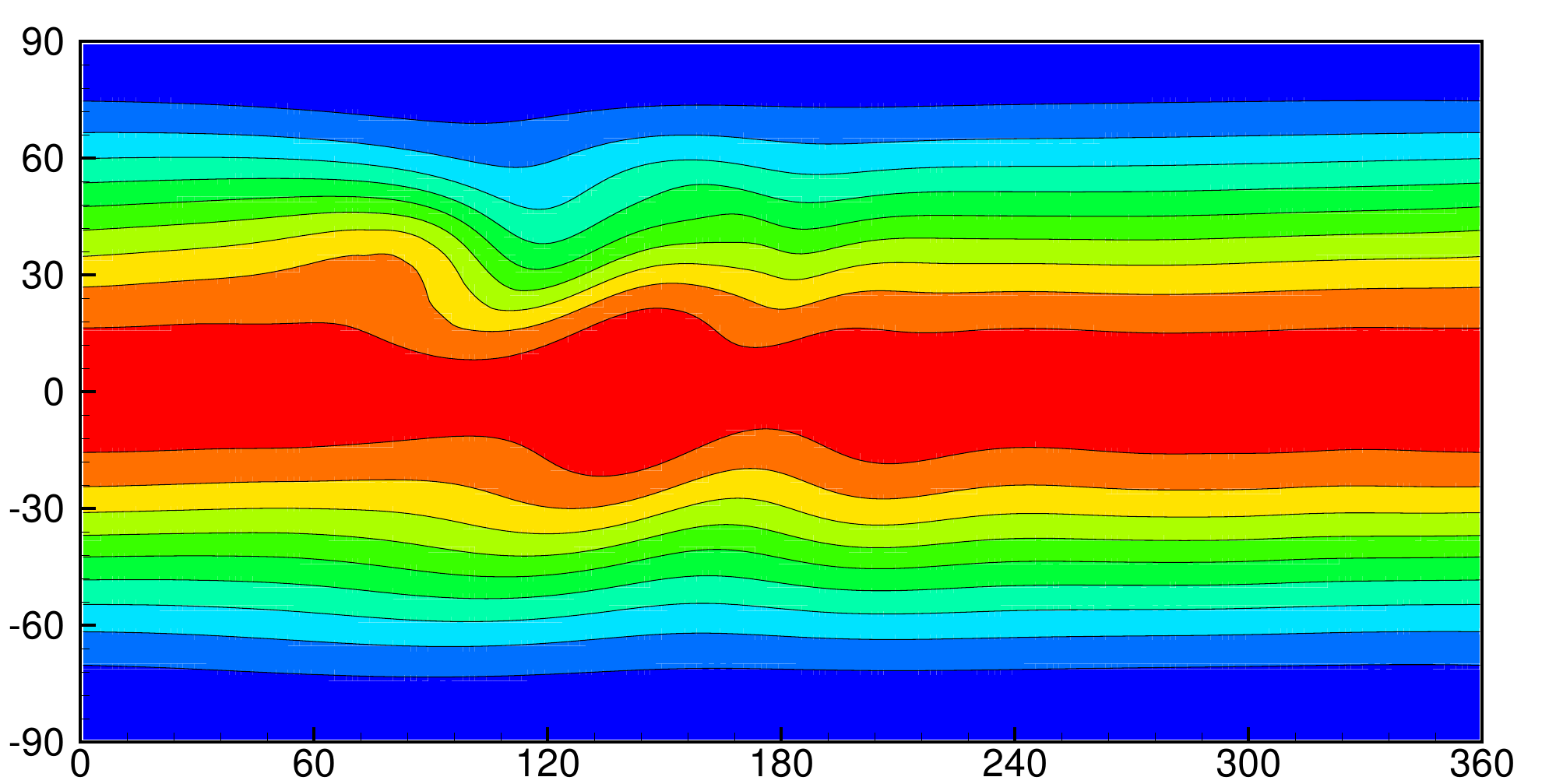}}
 \end{subfigure}
 \begin{subfigure}[700hPa temperature at day 5]
  { \includegraphics[width=0.48\textwidth]{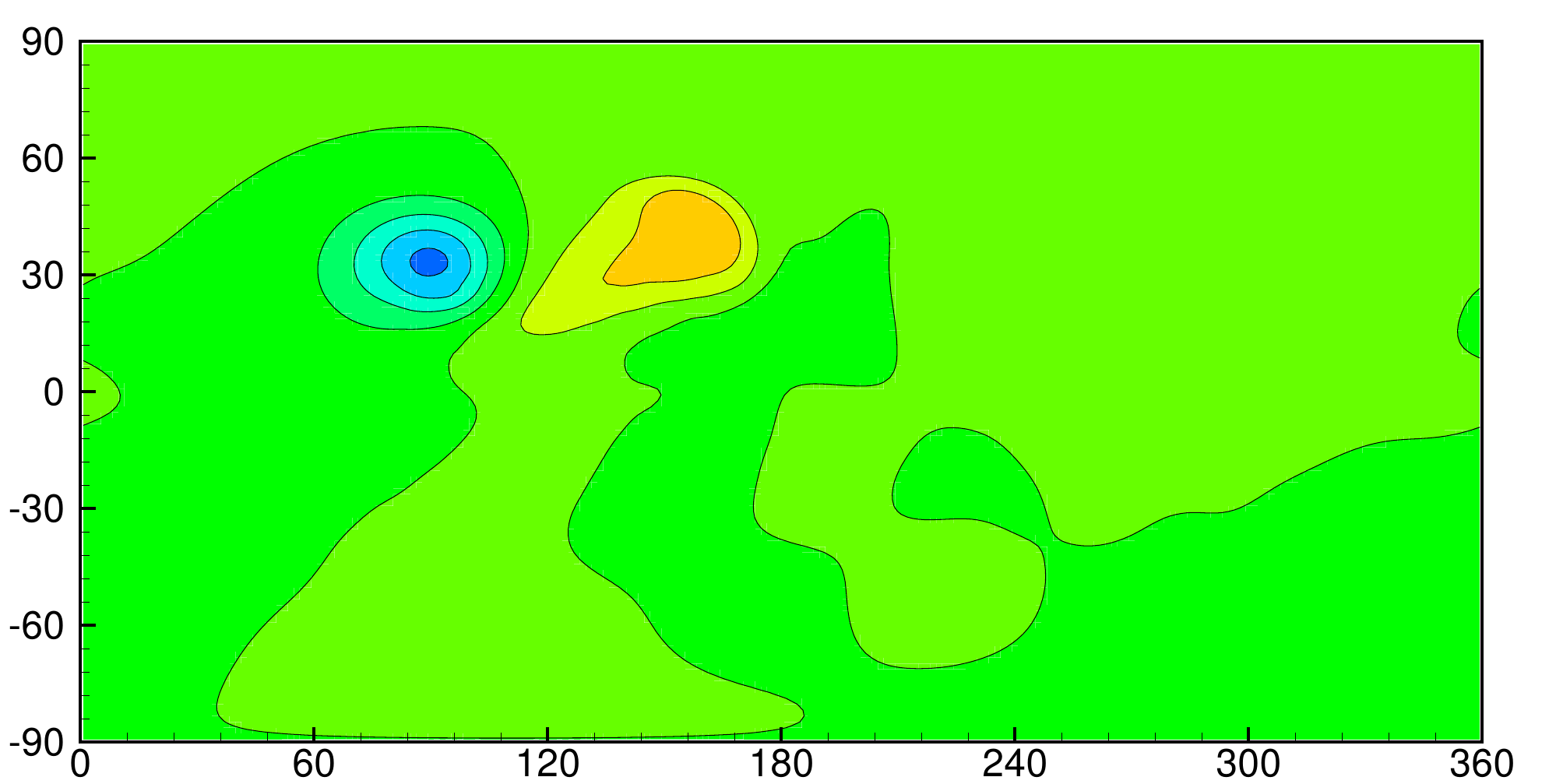}}
  \end{subfigure}
 \begin{subfigure}[700hPa geopotential height at day 15]
  {\includegraphics[width=0.48\textwidth]{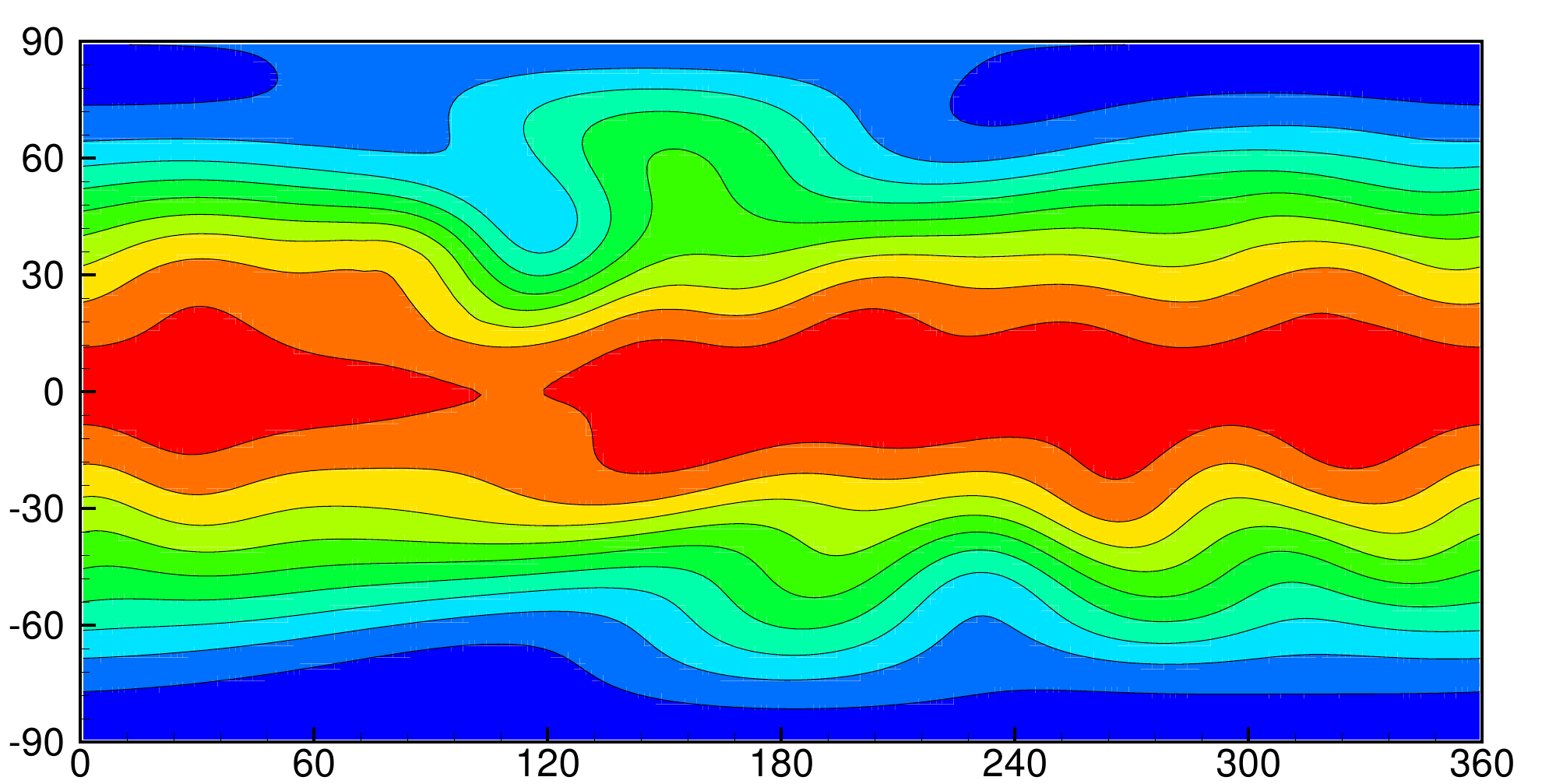}}
 \end{subfigure}
 \begin{subfigure}[700hPa temperature at day 15]
  { \includegraphics[width=0.48\textwidth]{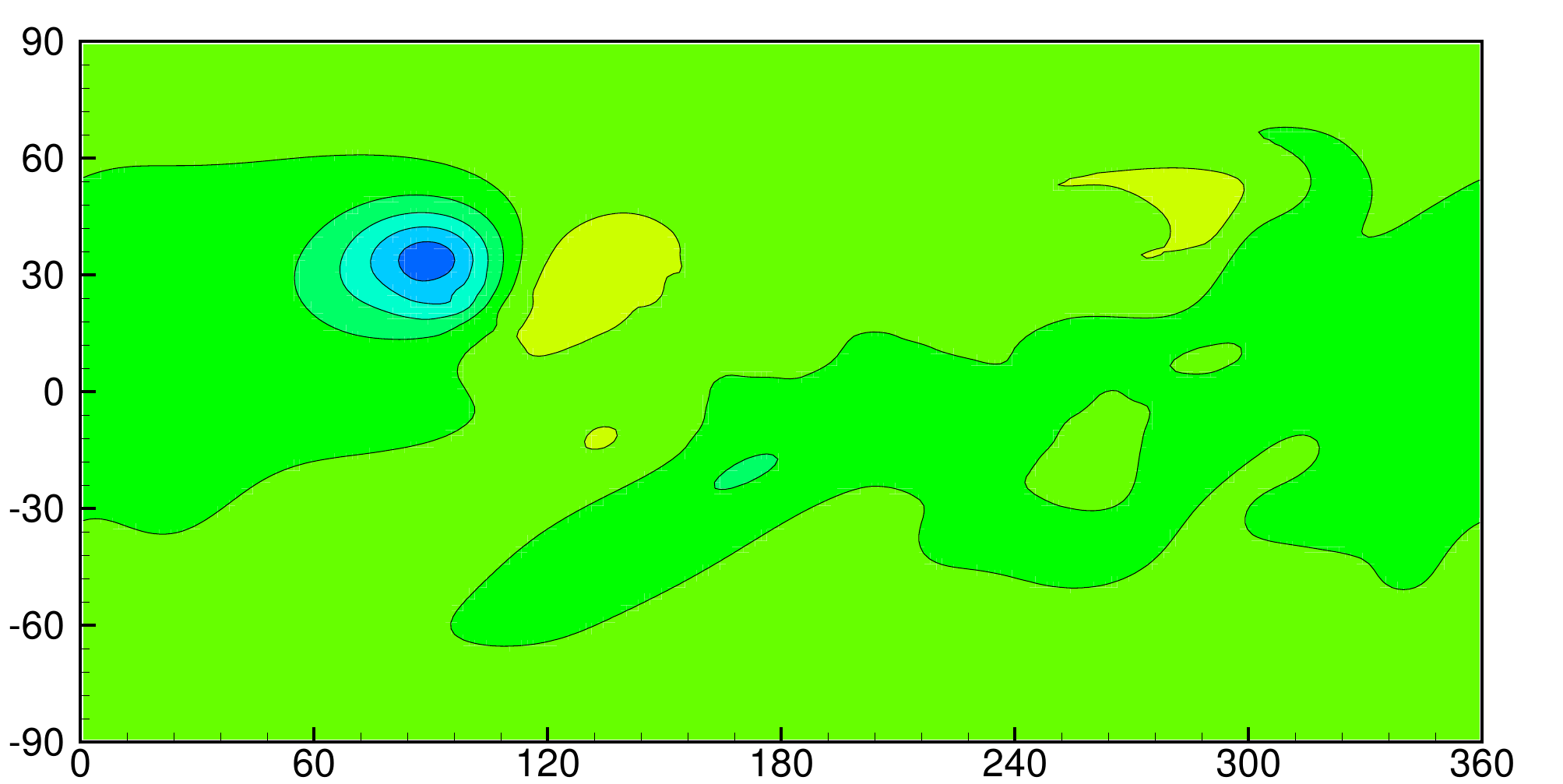}}
  \end{subfigure}
  \caption{Contour plots of 700hPa height and temperature of mountain-induced Rossby wave test on grid $30 \times 15$. Displayed contour lines vary within $\left[2500\mathrm{m},3300\mathrm{m}\right]$ with an interval of 100m for 700hPa height, vary within $\left[273\mathrm{K},300\mathrm{K}\right]$ with an interval of 3K for 700hPa temperature.}\label{MountainWave1}
\end{figure}

\begin{figure}[h]
 \centering
 \begin{subfigure}[Zonal velocity at day 5]
  { \includegraphics[width=0.48\textwidth]{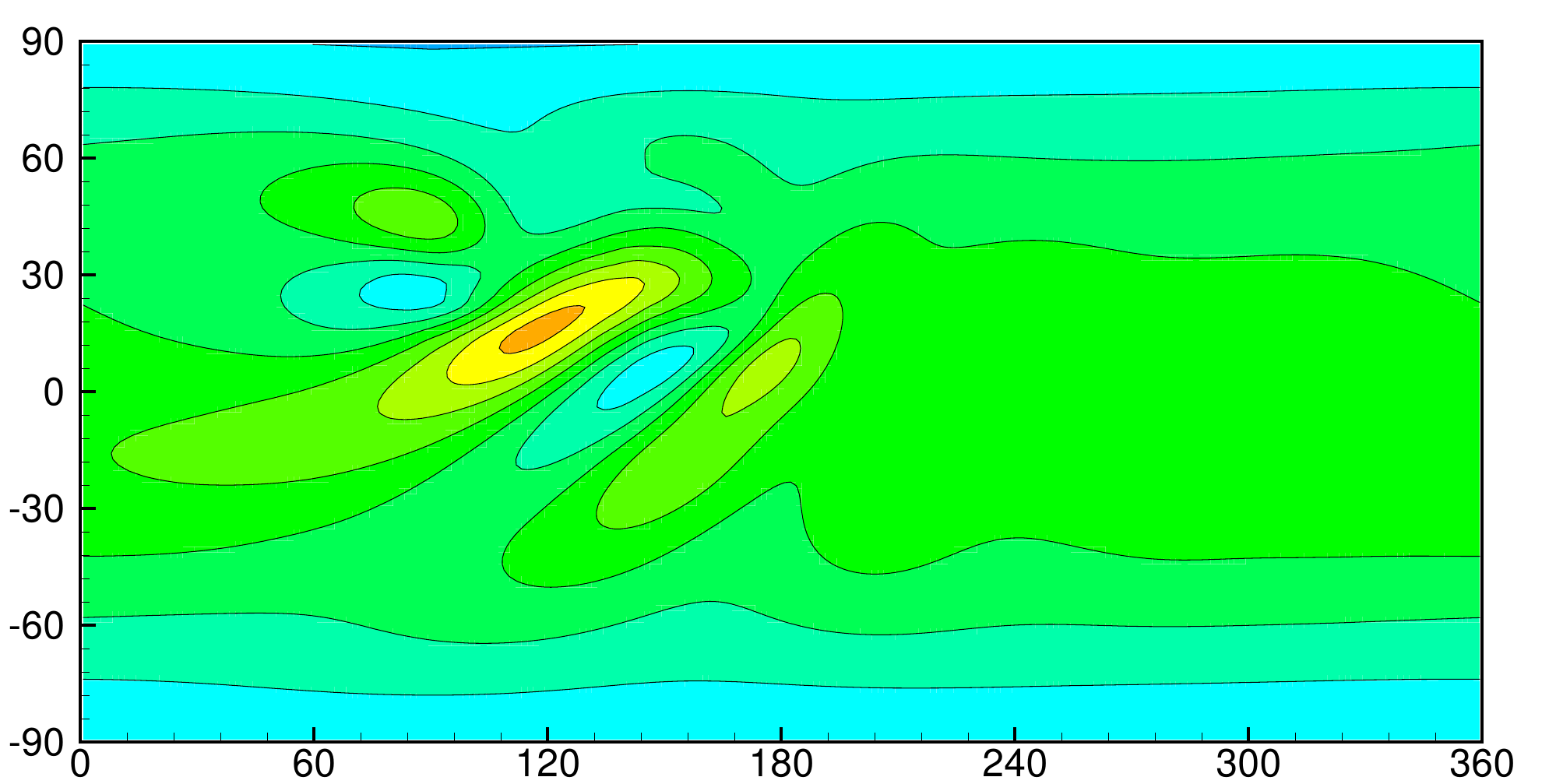}}
  \end{subfigure}
   \begin{subfigure}[meridional velocity at day 5]
   { \includegraphics[width=0.48\textwidth]{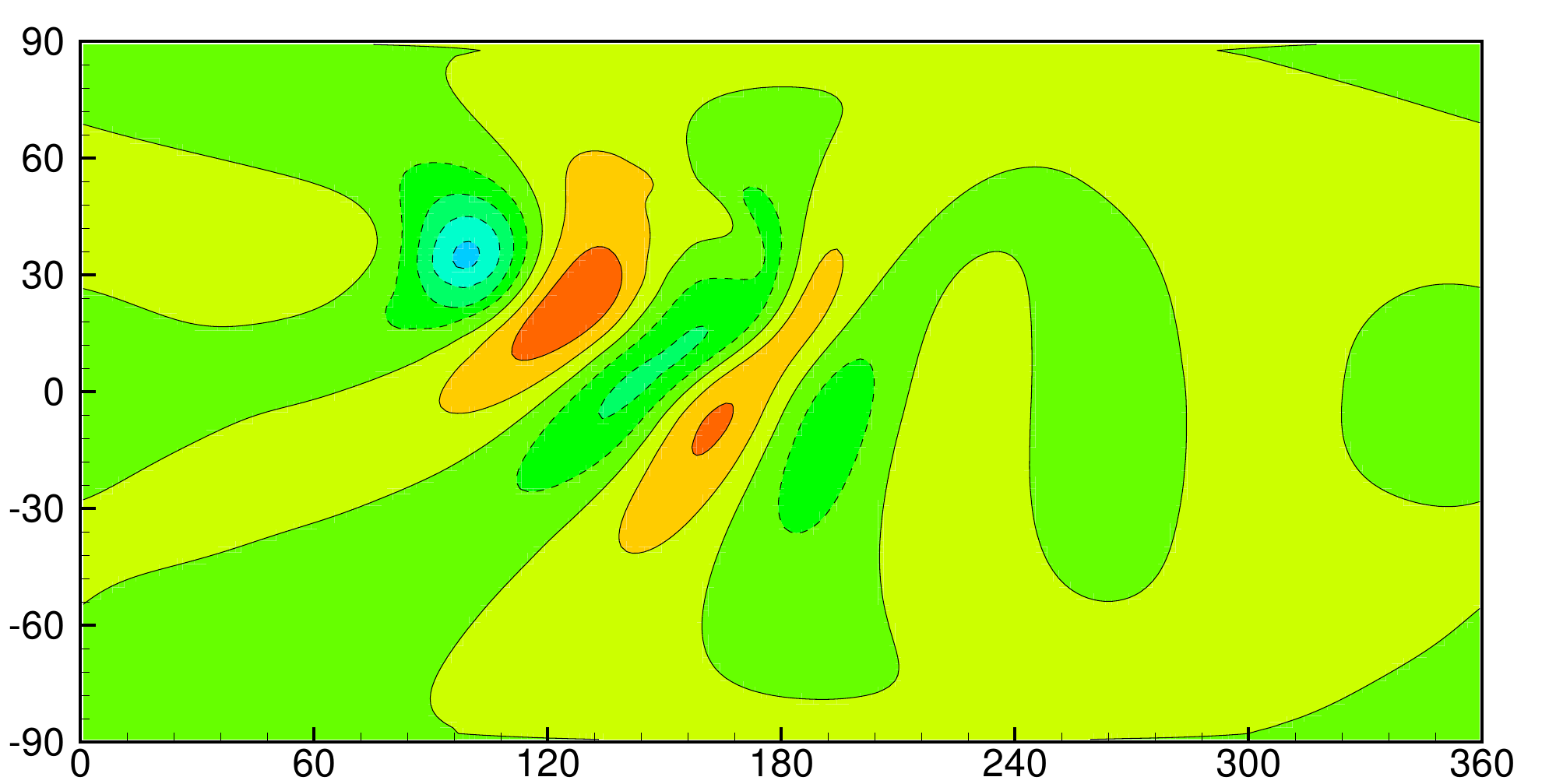}}
   \end{subfigure}
\begin{subfigure}[Zonal velocity at day 5]
  { \includegraphics[width=0.48\textwidth]{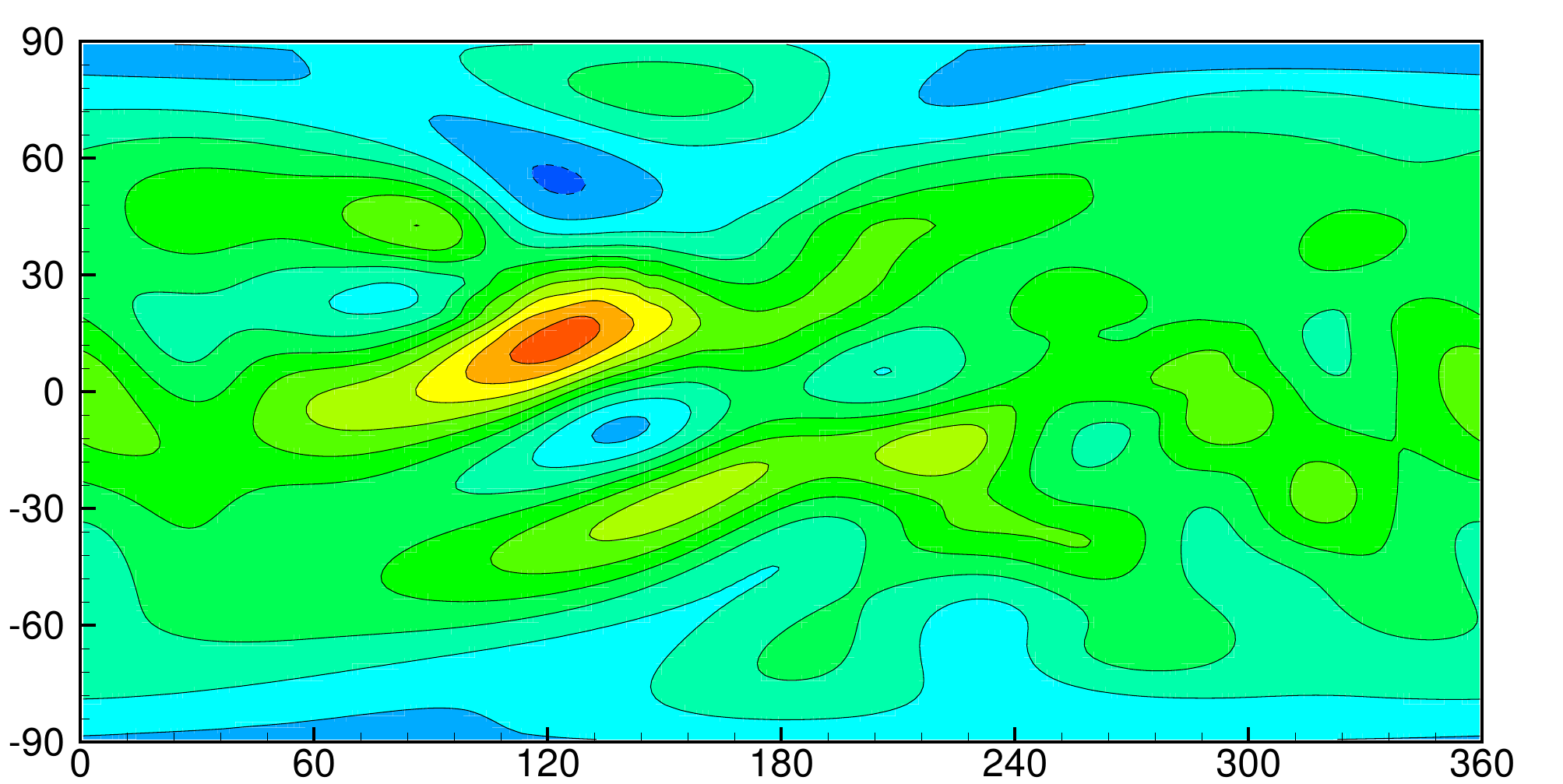}}
  \end{subfigure}
     \begin{subfigure}[meridional velocity at day 15]
   { \includegraphics[width=0.48\textwidth]{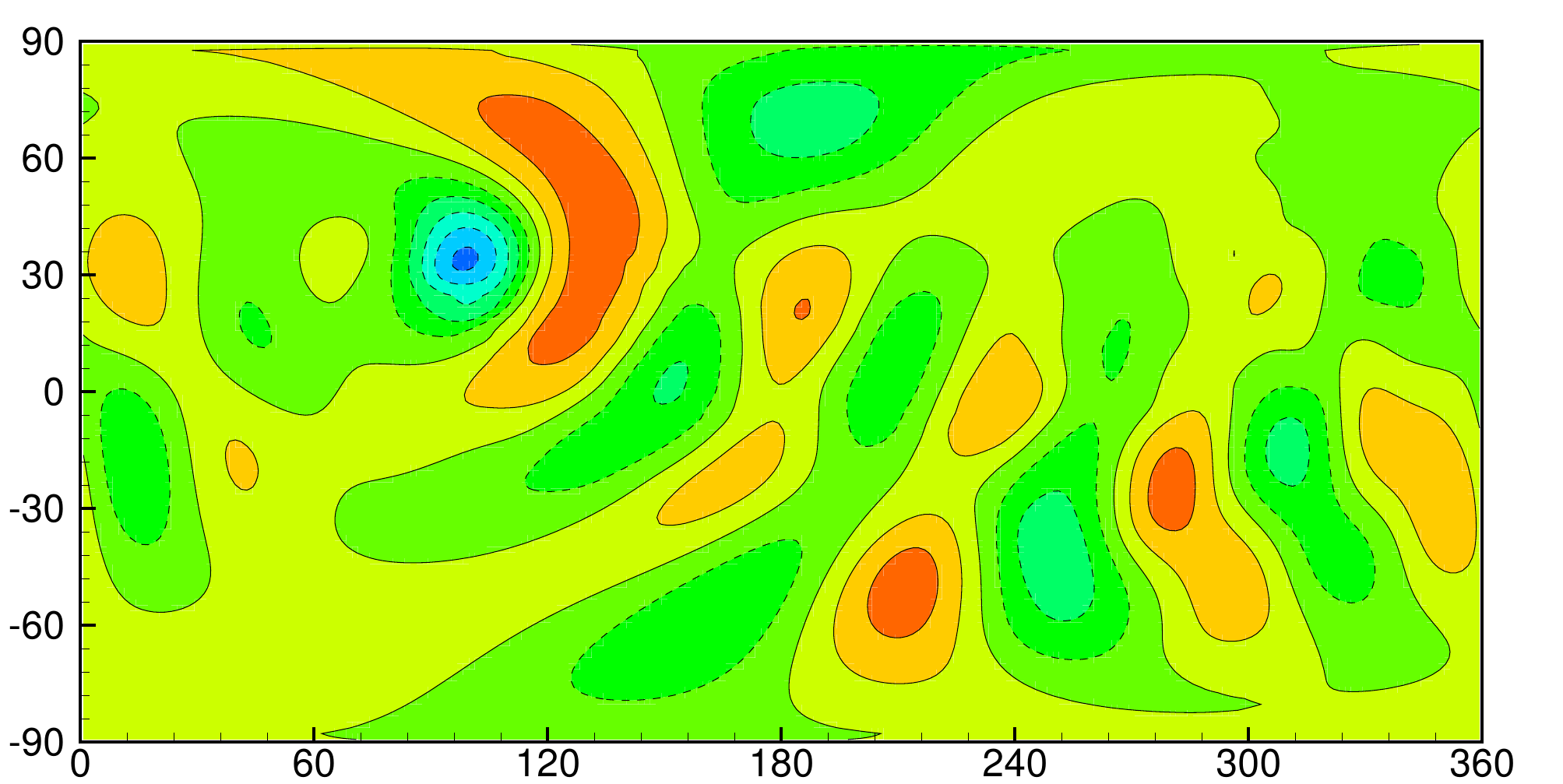}}
   \end{subfigure}
  \caption{Same as Fig.\ref{MountainWave1}, but for 700hPa velocity field. Displayed contour lines vary within $\left[-10\mathrm{m/s},45\mathrm{m/s}\right]$ with an interval of 5m/3 for zonal velocity, vary within $\left[-30\mathrm{m/s},15\mathrm{m/s}\right]$ with an interval of 5m/s for meridional velocity. The dashed lines are used for negative values.}\label{MountainWave2}
\end{figure}

\begin{figure}[h]
 \centering
  \begin{subfigure}[Surface pressure at day 7]
  { \includegraphics[width=0.48\textwidth]{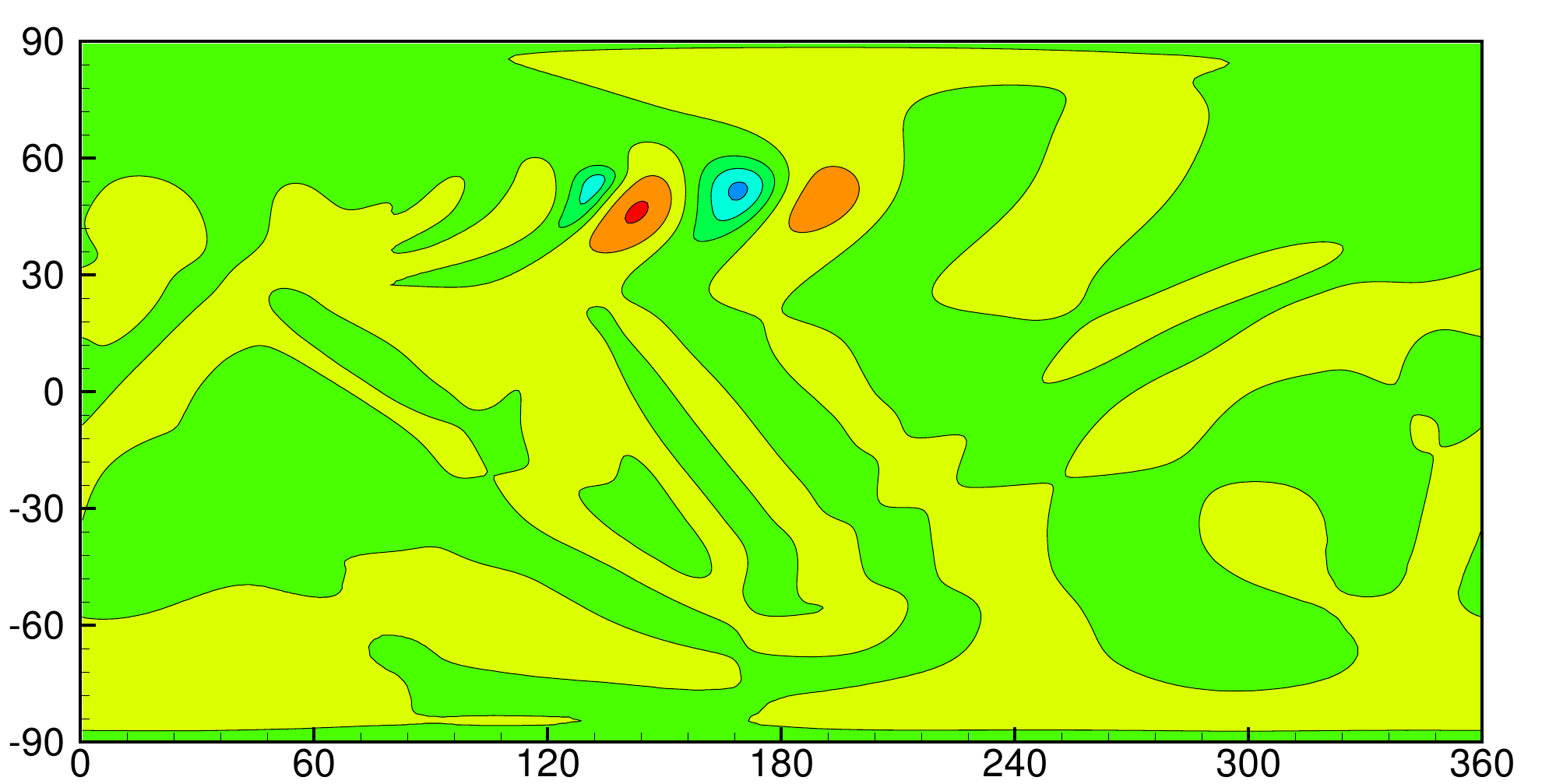}}
  \end{subfigure}
  \begin{subfigure}[Surface pressure at day 9]
  { \includegraphics[width=0.48\textwidth]{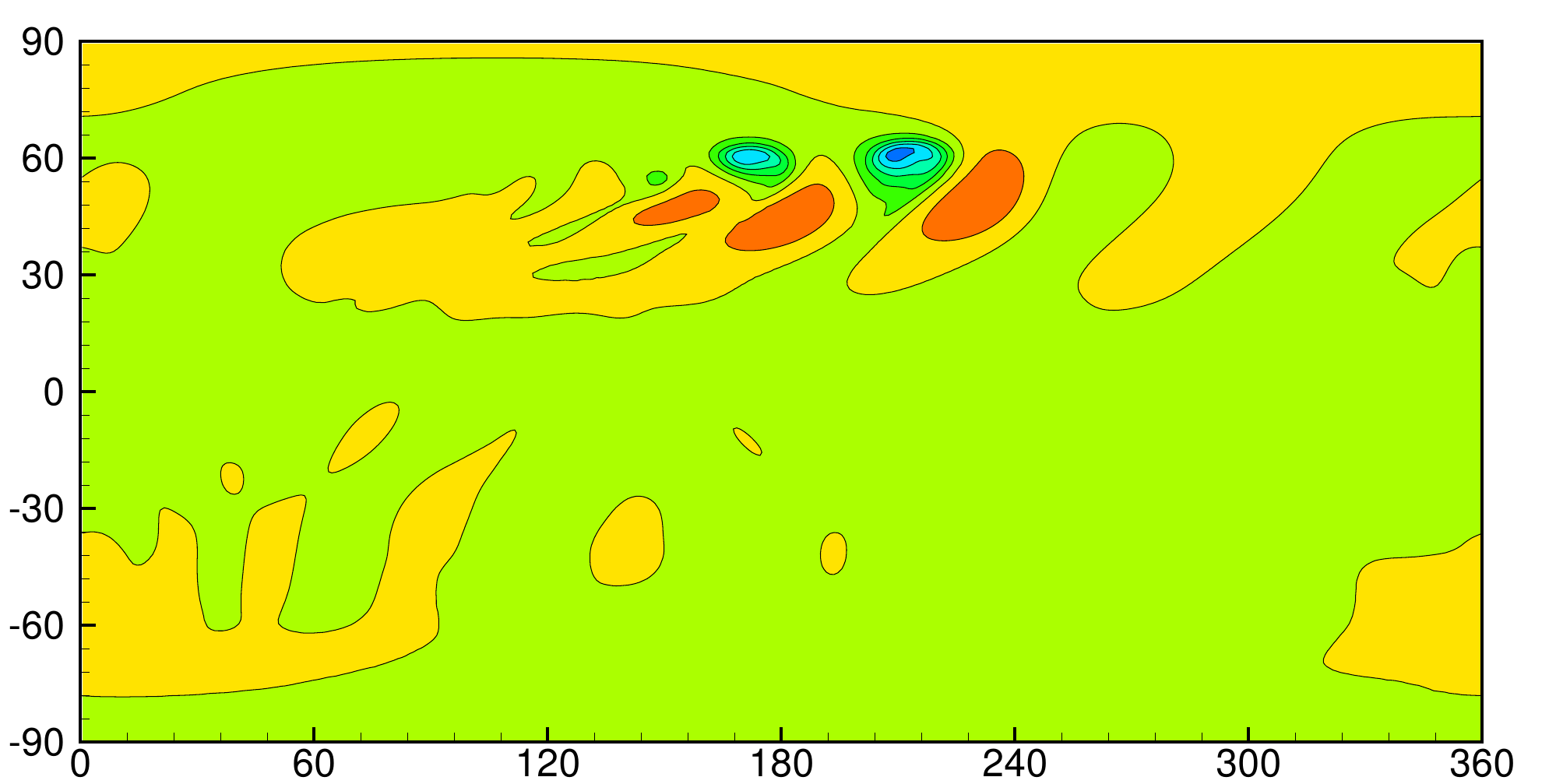}}\\
  \end{subfigure}
  \begin{subfigure}[850hPa temperature at day 7]
   { \includegraphics[width=0.48\textwidth]{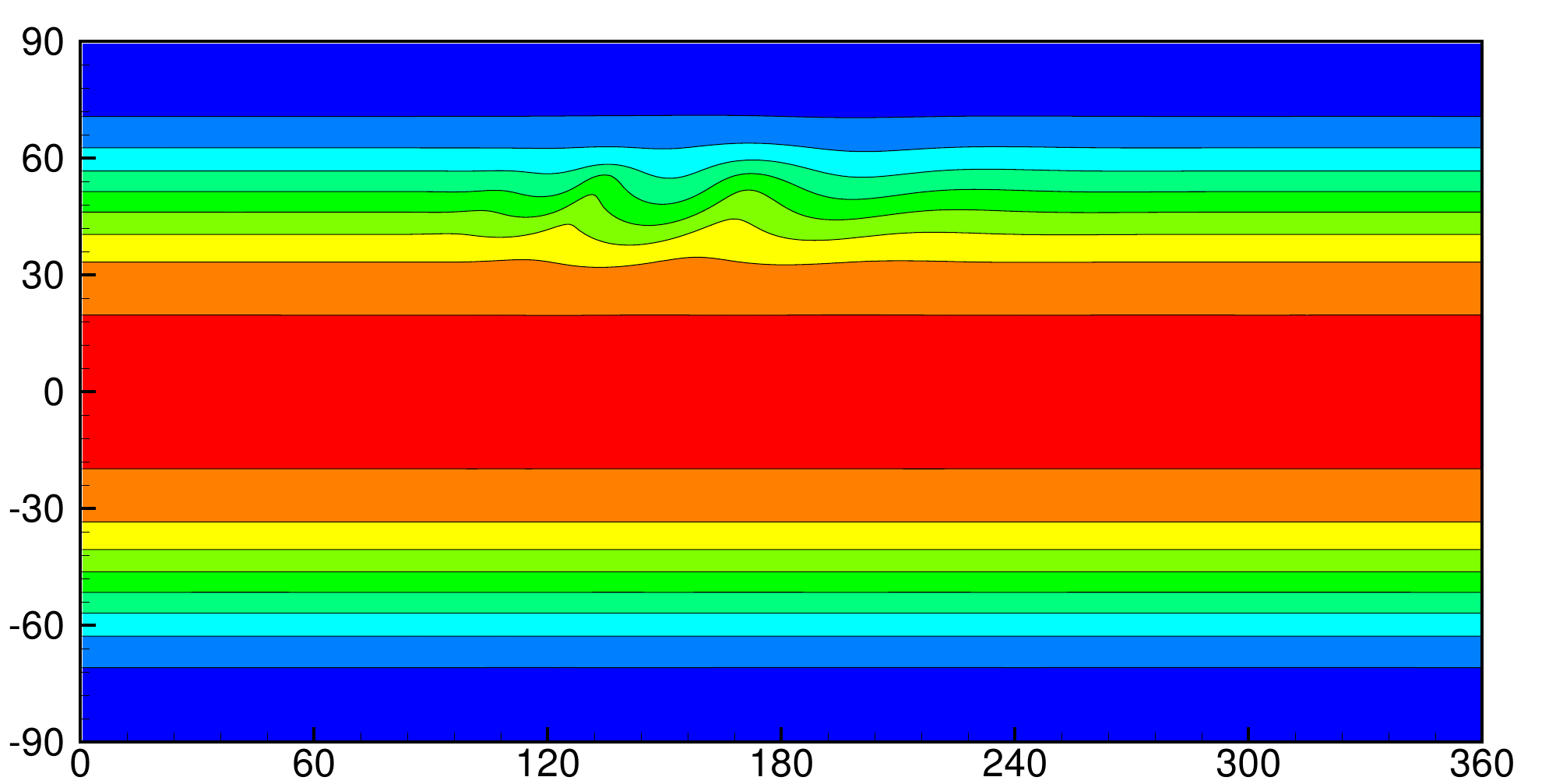}}
  \end{subfigure}
  \begin{subfigure}[850hPa temperature at day 9]
   { \includegraphics[width=0.48\textwidth]{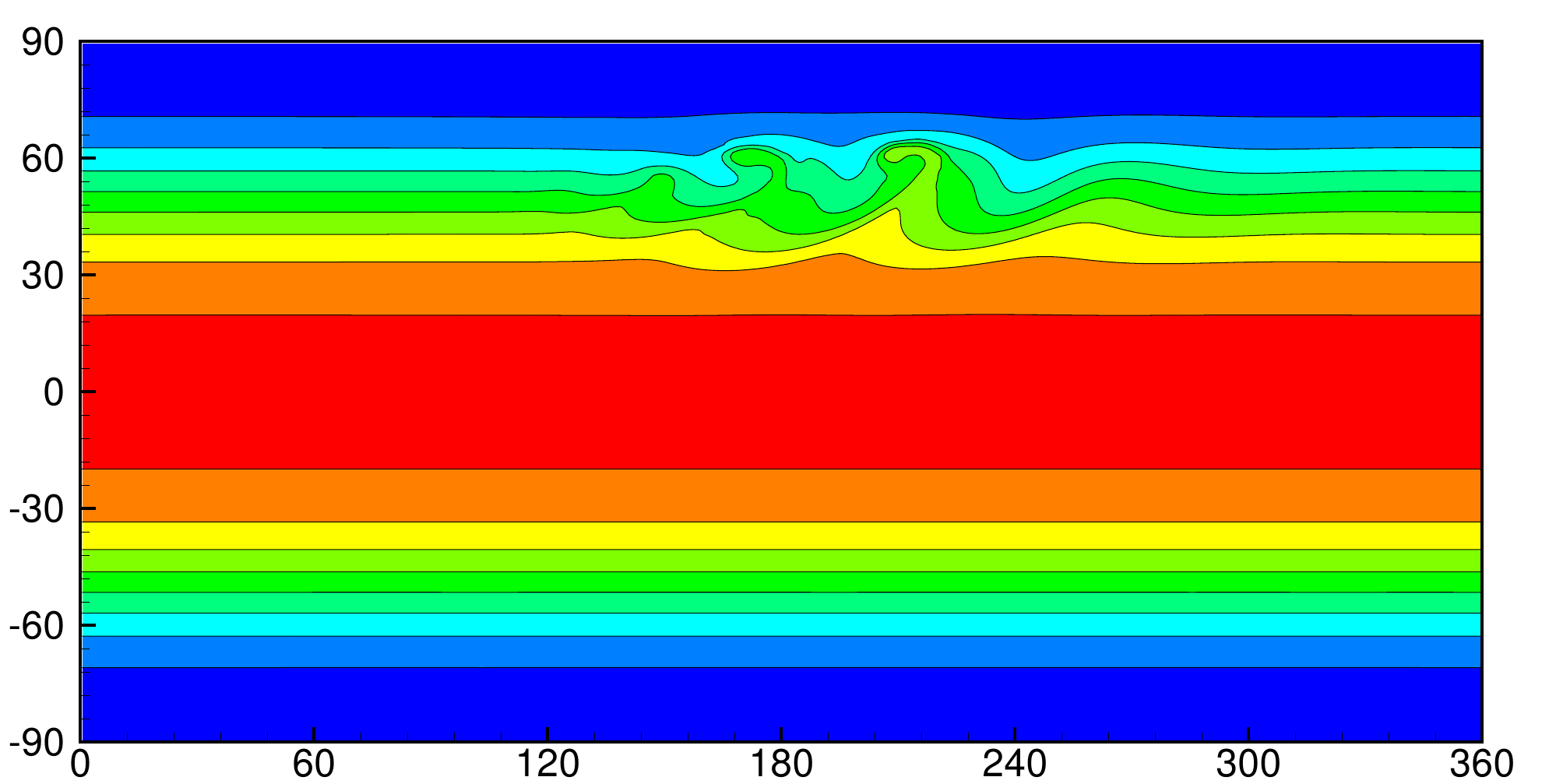}}\\
  \end{subfigure}
  \begin{subfigure}[850hPa relative vorticity at day 7]
  { \includegraphics[width=0.48\textwidth]{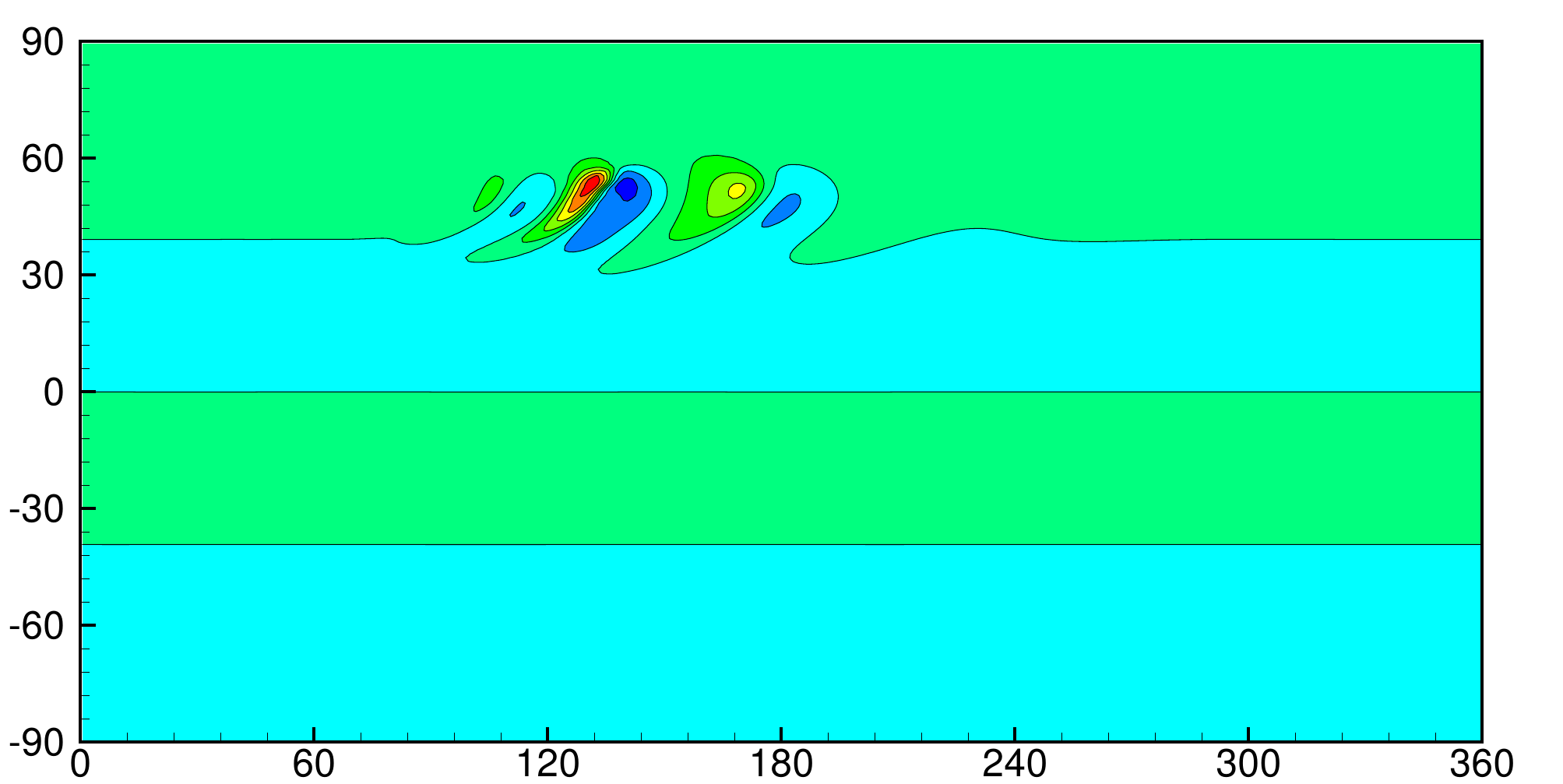}}
  \end{subfigure}
  \begin{subfigure}[850hPa relative vorticity at day 9]
  { \includegraphics[width=0.48\textwidth]{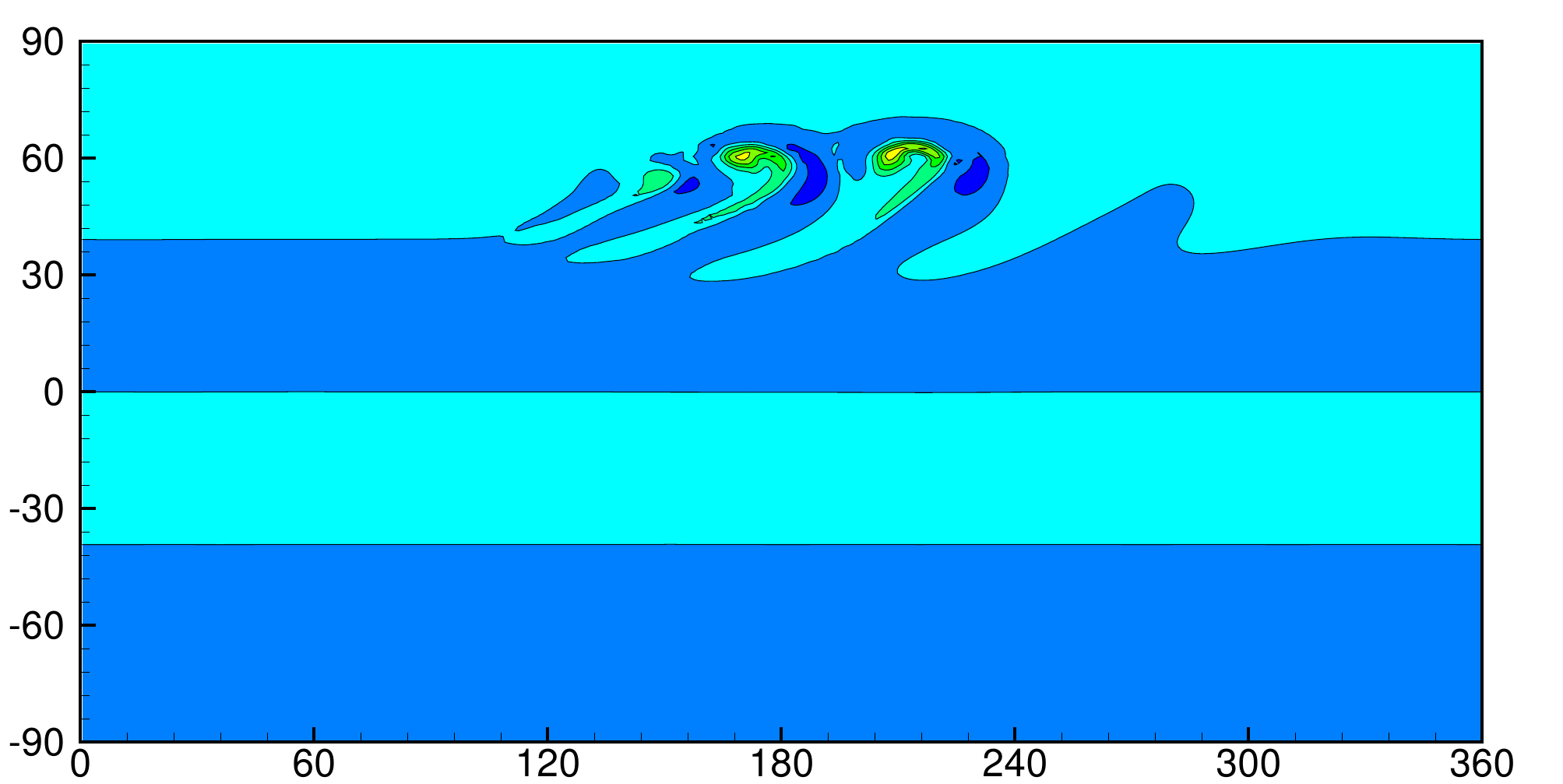}}
  \end{subfigure}
  \caption{Contour plots of numerical results of baroclinic wave test on grid $45 \times 15$. Displayed contour lines vary within $\left[984\mathrm{hPa},1008\mathrm{hPa}\right]$ with an interval of 4hPa for surface pressure at day 7, within $\left[940\mathrm{hPa},1020\mathrm{hPa}\right]$ with an interval of 10hPa for surface pressure at day 9, within $\left[230\mathrm{K},300\mathrm{K}\right]$ with an interval of 10K for 850hPa temperature, within $\left[-2\times10^5\mathrm{s}^{-1},5\times10^5\mathrm{s}^{-1}\right]$ with an interval of 1$\times10^5\mathrm{s}^{-1}$ for relative vorticity at day 7, within $\left[-5\times10^5\mathrm{s}^{-1},3\times10^4\mathrm{s}^{-1}\right]$ with an interval of 5$\times10^5\mathrm{s}^{-1}$ for relative vorticity at day 9. The dashed lines are used for negative values.}\label{BaroclinicWave}
\end{figure}

The numerical results are shown in Fig. \ref{RossbyWave} for 3D Rossby-Haurwitz wave case, Fig. \ref{GravityWave} for gravity wave , Figs. \ref{MountainWave1} and \ref{MountainWave2} for mountain-induced Rossby wave case and Fig. \ref{BaroclinicWave} for baroclinic wave case. All results agree well with the reference solutions given in \citep{jablonowski2008}.

\section{Summary}

A fourth-order nonhydrostatic dynamical core for global atmospheric model is proposed in this study by using multi-moment finite volume method. Though introducing at least two kinds of moments as model variables, the high order numerical schemes can be constructed over more compact spatial stencils compared with the traditional finite volume method. The multi-moment model is very flexible regarding the computational meshes with complex topologies. The proposed 3D dynamical core achieves fourth-order accuracy in spherical geometry with the application of the cubed-sphere grid. The benchmark tests proposed in \citep{jablonowski2008} are also tested. The numerical results agree well with reference solutions and reveal that the proposed model is capable of accurately reproducing these large-scale and quasi-hydrostaic atmospheric dynamics. The verification of numerical tests with more nonhydrostatic effect or for moist dynamical cores proposed in DCMIP 2012 \citep{DCMIP2012} and 2016 \cite{DCMIP2016} is being conducted.

\section*{Acknowledgments}

This work is supported by National Key Research and Development Program of China (grant nos. 2017YFC1501901 and 2017YFA0603901), National Natural Science Foundation of China (grant no. 41522504), and the 111 Project (B18040).

\section*{References}

\bibliography{ref}

\begin{thebibliography}{10}
\expandafter\ifx\csname url\endcsname\relax
  \def\url#1{\texttt{#1}}\fi
\expandafter\ifx\csname urlprefix\endcsname\relax\def\urlprefix{URL }\fi
\expandafter\ifx\csname href\endcsname\relax
  \def\href#1#2{#2} \def\path#1{#1}\fi

\bibitem{Xiao2004}
F.~{Xiao}, Unified formulation for compressible and incompressible flows by
  using multi-integrated moments {I}: one-dimensional inviscid compressible
  flow, J. Comput. Phys. 195~(2) (2004) 629--654.
\newblock \href {http://dx.doi.org/http://dx.doi.org/10.1016/j.jcp.2003.10.014}
  {\path{doi:http://dx.doi.org/10.1016/j.jcp.2003.10.014}}.

\bibitem{Xiao2006}
F.~{Xiao}, R.~{Akoh}, S.~{Ii}, {Unified formulation for compressible and
  incompressible flows by using multi-integrated moments {II}:
  multi-dimensional version for compressible and incompressible flows}, J.
  Comput. Phys. 213~(1) (2006) 31--56.
\newblock \href {http://dx.doi.org/10.1016/j.jcp.2005.08.002}
  {\path{doi:10.1016/j.jcp.2005.08.002}}.

\bibitem{ii2009}
S.~Ii, F.~Xiao, {High order multi-moment constrained finite volume method. Part
  I: Basic formulation}, Journal of Computational Physics 228~(10) (2009)
  3669--3707.

\bibitem{Staniforth2012}
A.~{Staniforth}, J.~{Thuburn}, Horizontal grids for global weather and climate
  prediction models :a review, Q.J.R.Meteorol.Soc. 138 (2012) 1--26.

\bibitem{chen2014}
C.~G. {Chen}, X.~L. {Li}, X.~S. {Shen}, F.~{Xiao}, Global shallow water models
  based on multi-moment constrained finite volume method and three
  quasi-uniform spherical grids, J. Comput. Phys. 271 (2014) 191--223.

\bibitem{chen2008}
C.~{Chen}, F.~{Xiao}, Shallow water model on cubed-sphere by multi-moment
  finite volume method, J. Comput. phys. 227 (2008) 5019--5044.

\bibitem{mcore2012}
P.~A. {Ullrich}, C.~{Jablonowski}, {MCore: A non-hydrostatic atmospheric
  dynamical core utilizing high-order finite-volume methods}, J. Comput. Phys.
  231 (2012) 5078--5108.

\bibitem{clark1977}
T.~L. Clark, A small-scale dynamics model using a terrain-following coordinate
  transformation, J. Comput. Phys. 24 (1977) 186--215.

\bibitem{schar2002}
C.~{Sch\"{a}r}, D.~{Leuenberger}, O.~{Fuhrer}, D.~{L\"{u}thi}, C.~{Girard}, A
  new terrain-following vertical coordinate formulation for atmospheric
  prediction models, Mon. Wea. Rev. 130 (2002) 2459–2480.

\bibitem{Nair2005a}
R.~D. {Nair}, S.~J. {Thomas}, R.~D. {Loft}, {A discontinuous Galerkin transport
  scheme on the cubed sphere}, Mon. Wea. Rev. 133~(4) (2005) 827--841.

\bibitem{Nair2005b}
R.~D. {Nair}, S.~J. {Thomas}, R.~D. {Loft}, {A discontinuous Galerkin global
  shallow water model}, Mon. Wea. Rev. 133~(4) (2005) 876--887.

\bibitem{chen2011}
C.~{Chen}, F.~{Xiao}, X.~{Li}, An adaptive multimoment global model on a cubed
  sphere, Mon. Wea. Rev. 139 (2011) 523--548.

\bibitem{chen2015}
C.~G. {Chen}, X.~L. {Li}, X.~S. {Shen}, F.~{Xiao}, A high-order conservative
  collocation scheme and its application to global shallow-water equations,
  Geosci. Model Dev. 8 (2015) 221--233.

\bibitem{BGS}
X.~Deng, Z.~Sun, B.~Xie, K.~Yokoi, C.~Chen, F.~Xiao, A non-oscillatory
  multi-moment finite volume scheme with boundary gradient switching, Journal
  of Scientific Computing 72~(3) (2017) 1146--1168.
\newblock \href {http://dx.doi.org/10.1007/s10915-017-0392-0}
  {\path{doi:10.1007/s10915-017-0392-0}}.

\bibitem{WENO}
Z.~Sun, H.~Teng, F.~Xiao, {A Slope Constrained 4th Order Multi-Moment Finite
  Volume Method with WENO Limiter}, Communications in Computational Physics
  18~(4) (2015) 901--930.
\newblock \href {http://dx.doi.org/10.4208/cicp.081214.250515s}
  {\path{doi:10.4208/cicp.081214.250515s}}.

\bibitem{Paul2010}
P.~A. {Ullrich}, C.~{Jablonowski}, B.~{van Leer}, High-order finite-volume
  methods for the shallow-water equations on the sphere, J. Comput. Phys. 229
  (2010) 6104--6134.

\bibitem{Durran1983}
D.~L. {Durran}, J.~B. {Klemp}, A compressible model for the simulation of moist
  mountain waves, Mon. Wea. Rev. 111 (1983) 2341--2361.

\bibitem{weller2013}
H.~{Weller}, S.-J. {Lock}, N.~{Wood}, {Runge-Kutta IMEX schemes for the
  Horizontally Explicit/Vertically implicit (HEVI) solution of wave equations},
  J. Comput. Phys. 252 (2013) 365--381.

\bibitem{Gardner2018}
D.~J. {Gardner}, J.~E. {Guerra}, F.~P. {Hamon}, D.~R. {Reynolds}, P.~A.
  {Ullrich}, C.~S. {Woodward}, {Implicit--explicit (IMEX) Runge--Kutta methods
  for non-hydrostatic atmospheric models}, Geosci. Model Dev. 11 (2018)
  1497--1515.

\bibitem{ascher1997}
U.~M. {Ascher}, S.~J. {Ruuth}, R.~J. {Spiteri}, {Implicit-explicit Runge-Kutta
  methods for time-dependent partial different equaitons}, Applied Numerical
  Mathematics 25 (1997) 151--167.

\bibitem{jablonowski2008}
C.~{Jablonowski}, P.~{Lauritzen}, R.~{Nair}, M.~{Taylor}, {idealized test cases
  for the dynamical cores of Atmospheric Genercal Circulation Models: A
  proposal for the NCAR ASP 2008 summer colloquium}, Tech. rep. (2008).

\bibitem{DCMIP2012}
P.~A. {Ullrich}, C.~{Jablonowski}, P.~H. {Lauritzen}, R.~D. {Nair}, M.~A.
  {Taylor}, {Dynamical core model intercomparison project (DCMIP) test case
  document}, Tech. rep., {DCMIP summer school} (2012).

\bibitem{DCMIP2016}
P.~A. {Ullrich}, C.~{Jablonowski}, K.~A. {Reed}, C.~{Zarzycki}, P.~H.
  {Lauritzen}, R.~D. {Nair}, J.~{Kent}, A.~{Verlet-Banide}, {Dynamical core
  model intercomparison project (DCMIP2016) test case document}, Tech. rep.,
  {DCMIP summer school} (2016).

\bibitem{ARPS}
ARPS User's Guide (Version 4.0), http://www.caps.ou.edu/ARPS/arpsdoc.html.

\end{thebibliography}

\begin{appendix}

\section{Non-uniform vertical grid}\label{verticalgrid}

Here we briefly introduce the non-uniform vertical coordinate currently adopted in this study, which is designed to refine the grid near the surface to better represent the surface topography. As described in section 2, the uniform vertical coordinate in computational space is denoted by $\zeta\in\left[0,r_t\right]$ with grid spacing of $\Delta \zeta=\frac{r_t}{N_v}$, the corresponding non-uniform one is obtained by using transformation $\hat{\zeta}=\mathcal{T}\left(\zeta\right)$.

For the non-uniform coordinate in the computational space, the smallest grid spacing of $\hat{\zeta}$ is $\Delta \hat{\zeta}_{\min}$ and the largest one is $\Delta \hat{\zeta}_{\max}$. We define the parameters $\delta_1=\frac{\Delta \hat{\zeta}_{\min}}{\Delta \zeta}$, $\delta_2=\frac{\Delta \hat{\zeta}_{\max}}{\Delta \zeta}$ and $\delta_m=\frac{\delta_1+\delta_2}{2}$.

The non-uniform coordinate is derived by choosing $\mathcal{T}_\zeta\left(\zeta\right)$ as,
\begin{equation}
\mathcal{T}_\zeta\left(\zeta\right)=\left\{
\begin{array}{l}
\delta_1,\ \mathrm{if}\ \zeta<\zeta_1\\
\delta_m+\frac{1}{2}\delta_{12}\sin\left(\frac{\zeta-\zeta_m}{\zeta_{12}}\pi\right),\ \mathrm{if}\ \zeta_1\leq \zeta \leq \zeta_2\\
\delta_2,\ \mathrm{otherwise}
\end{array}
\right.,\label{vgrid1}
\end{equation}
where $\delta_{12}=\delta_2-\delta_1$, $\zeta_{12}=\zeta_2-\zeta_1$ and $\zeta_m=\frac{\zeta_1+\zeta_2}{2}$.

\begin{figure}[h]
 \centering
 \begin{subfigure}[$\mathcal{T}_\zeta(\zeta)$]
  { \includegraphics[width=0.48\textwidth]{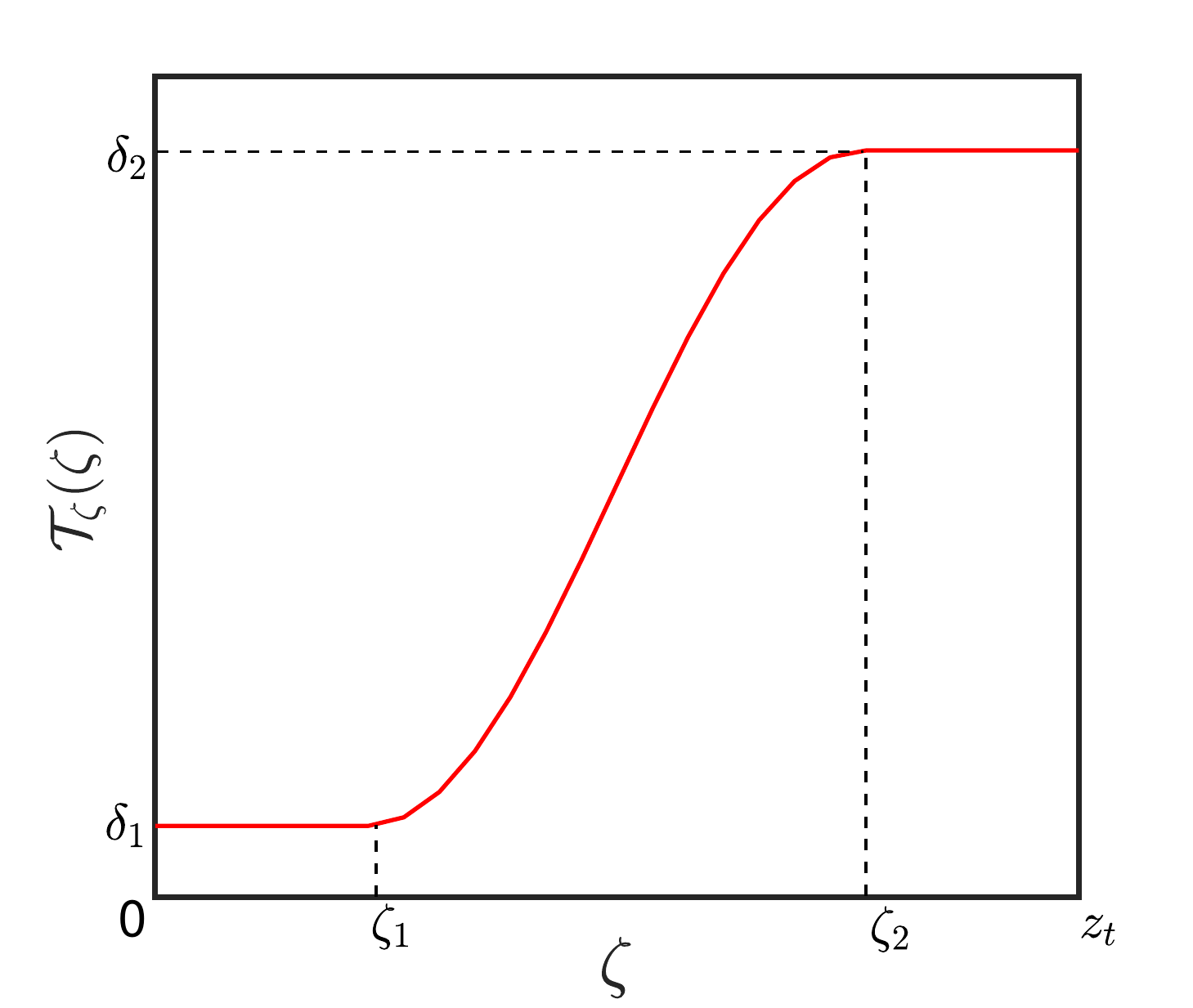}}
  \end{subfigure}
   \begin{subfigure}[$\hat{\zeta}$]
   { \includegraphics[width=0.48\textwidth]{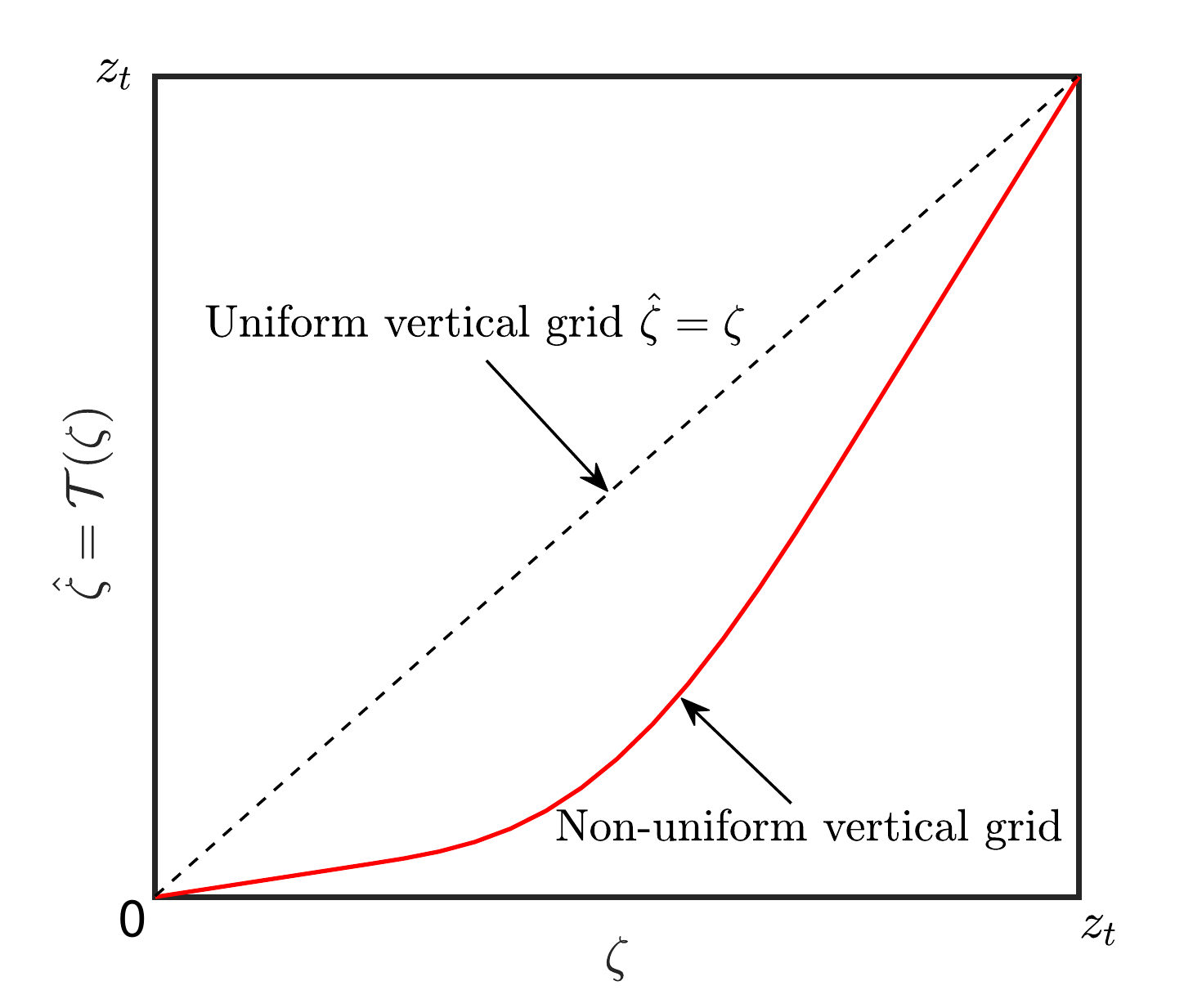}}
   \end{subfigure}
  \caption{Illustration of non-uniform vertical coordinate.\label{VerticalGridFig}}
\end{figure}

As shown in Fig.\ref{VerticalGridFig}, several grid points is equidistantly arranged near the surface ($0 < \zeta < \zeta_1$) and the model top ($\zeta_2<\zeta<z_t$). Near the surface the finest resolution $\Delta \hat{\zeta}_{\min}$ is used, whereas the coarsest one $\Delta \hat{\zeta}_{\max}$ is set near the top. Within $\left[\zeta_1,\zeta_2\right]$, the derivative of transformation $\mathcal{T}$ increases from $\delta_1$ to $\delta_2$ and has a shape of the sinusoid in this study. Similar arrangement of the non-uniform vertical coordinate is designed in ARPS (Advanced Regional Predication System) \cite{ARPS}.

By integrating Eq.\eqref{vgrid1}, we can decide the location of the non-uniform vertical coordinate as
\begin{equation}
\mathcal{T}\left(\zeta\right)=\left\{
\begin{array}{l}
\delta_1 \zeta,\ \mathrm{if}\ \zeta<\zeta_1\\
\delta_m \zeta - \frac{1}{2\pi}\delta_{12}\zeta_{12}\cos\left(\frac{\zeta-\zeta_m}{\zeta_{12}}\pi\right)-\frac{1}{2}\delta_{12}\zeta_1,\ \mathrm{if}\ \zeta_1\leq \zeta \leq \zeta_2\\
\delta_2 \zeta -\delta_{12}\zeta_m,\ \mathrm{otherwise}
\end{array}
\right.,\label{vgrid2}
\end{equation}
subjected to the relation
\begin{equation}
r_t=\mathcal{T}\left(r_t\right)=\delta_2r_t-\zeta_m\delta_{12}.\label{vgrid3}
\end{equation}

In two test cases with topography, we choose the non-uniform grid with the parameters shown in Table \ref{vgridtable} and $\zeta_2$ is determined using relation \eqref{vgrid3}.

\begin{table*}[ht]
{
\caption{Parameters for constructing vertical grid.}\label{vgridtable}
\begin{tabularx}{\textwidth}{l@{\extracolsep{\fill}}ccccccc}
\hline\hline
&  Case                   & model top &  $N_v $ &  $\Delta\hat{\zeta}_{\min}$   & $\Delta\hat{\zeta}_{\max}$  & $\zeta_1$  \\ \hline
& Mountain wave           & 30km      & 15      & 400m                      & 2800m                   & 1200m     \\
& Baroclinic wave         & 44km      & 15      & 120m                      & 4800m                   & 120m   \\ \hline
\end{tabularx}}
\end{table*}

\end{appendix}

\end{document}